%% file: neurips_2026.tex
\documentclass{article}

\PassOptionsToPackage{numbers, compress}{natbib}

\usepackage[main,final,nonatbib]{neurips_2026}

\usepackage[utf8]{inputenc} 
\usepackage[T1]{fontenc}    
\usepackage{CJKutf8}        
\usepackage[hypertexnames=false]{hyperref}       
\usepackage{url}            
\usepackage{booktabs}       
\usepackage{amsfonts}       
\usepackage{amsmath}        
\usepackage{nicefrac}       
\usepackage{microtype}      
\usepackage{graphicx}       
\usepackage{caption}        
\usepackage{multirow}       
\usepackage[table]{xcolor}  
\usepackage{algorithm}      
\usepackage{algpseudocode}  
\usepackage{placeins}       
\usepackage{tcolorbox}      
\tcbuselibrary{breakable}
\usepackage{listings}       
\usepackage{enumitem}       
\definecolor{customblue}{RGB}{33, 102, 172}
\definecolor{customblue_back}{RGB}{239, 246, 255}
\definecolor{codekeyword}{RGB}{0, 92, 175}
\definecolor{codestring}{RGB}{163, 21, 21}
\definecolor{codecomment}{RGB}{0, 128, 0}
\definecolor{codenumber}{RGB}{128, 128, 128}
\definecolor{SlateBlue}{RGB}{90, 122, 168}
\definecolor{gainorange}{RGB}{255, 102, 51}
\definecolor{methodhighlight}{RGB}{255, 244, 230}
\lstdefinestyle{prismpython}{
  language=Python,
  basicstyle=\ttfamily\scriptsize,
  keywordstyle=\color{codekeyword}\bfseries,
  stringstyle=\color{codestring},
  commentstyle=\color{codecomment}\itshape,
  breaklines=true,
  breakatwhitespace=false,
  columns=fullflexible,
  keepspaces=true,
  showstringspaces=false,
  upquote=true,
  frame=none,
  tabsize=4,
  morekeywords={self,async,await,True,False,None},
  emph={__init__,__call__,_fill_node,_ask_text,_parse_subtasks,_exec_code_impl,exec_code},
  emphstyle=\color{customblue}\bfseries,
}
\lstdefinelanguage{json}{
  basicstyle=\ttfamily\scriptsize,
  showstringspaces=false,
  breaklines=true,
  breakatwhitespace=false,
  columns=fullflexible,
  keepspaces=true,
  upquote=true,
  literate=
   *{:}{{{\color{codekeyword}{:}}}}{1}
    {,}{{{\color{codekeyword}{,}}}}{1}
    {\{}{{{\color{customblue}{\{}}}}{1}
    {\}}{{{\color{customblue}{\}}}}}{1}
    {[}{{{\color{customblue}{[}}}}{1}
    {]}{{{\color{customblue}{]}}}}{1}
    {0}{{{\color{codenumber}0}}}{1}
    {1}{{{\color{codenumber}1}}}{1}
    {2}{{{\color{codenumber}2}}}{1}
    {3}{{{\color{codenumber}3}}}{1}
    {4}{{{\color{codenumber}4}}}{1}
    {5}{{{\color{codenumber}5}}}{1}
    {6}{{{\color{codenumber}6}}}{1}
    {7}{{{\color{codenumber}7}}}{1}
    {8}{{{\color{codenumber}8}}}{1}
    {9}{{{\color{codenumber}9}}}{1},
}
\lstdefinestyle{prismjson}{
  language=json,
  basicstyle=\ttfamily\scriptsize,
  stringstyle=\color{codestring},
  breaklines=true,
  breakatwhitespace=false,
  columns=fullflexible,
  keepspaces=true,
  showstringspaces=false,
  upquote=true,
  frame=none,
  tabsize=2,
}
\newcommand{\method}{FlowMAS}
\newcommand{\best}[1]{\textbf{#1}}
\newcommand{\second}[1]{\underline{#1}}
\newcommand{\groupsep}{\midrule}

\makeatletter
\let\prismincludegraphics\includegraphics
\renewcommand{\includegraphics}[2][]{%
  \IfFileExists{#2}{%
    \prismincludegraphics[#1]{#2}%
  }{%
    \fbox{\parbox[c][3cm][c]{0.9\linewidth}{\centering\footnotesize Missing figure\\{\ttfamily\detokenize{#2}}}}%
  }%
}
\makeatother

\title{FlowMAS: Learning Multi-Agent Workflow Topology via Information-guided Generative Flow Network}

\author{
  \hspace*{\dimexpr -35pt}
  \begin{tabular}{@{}c@{}}
    Haitao Wang$^{ab}$\thanks{Equal contribution} ,
    Chenjing Liang$^{ab}$\footnotemark[1] ,
    Haipeng Zhang$^{ab}$, Jiawei Hu$^{ab}$, Sicheng Wang$^{ab}$, \\ 
    \textbf{Songzhu Mei$^{c}$}, \textbf{Chenglu Wen$^{ab}$}, \textbf{Siqi Shen$^{ab}$}\thanks{Corresponding author}, \textbf{Cheng Wang$^{ab}$} \\
    \normalfont
    $^a$Fujian Key Laboratory of Urban Intelligent Sensing and Computing, \\
    \normalfont
    Xiamen University, Xiamen, China \\
    \normalfont
    $^b$Key Laboratory of Multimedia Trusted Perception and Efficient Computing, \\
    \normalfont
    Ministry of Education of China, Xiamen University, Xiamen, China \\
    \normalfont
    $^c$School of Computer, National University of Defense Technology, China \\
    \texttt{\{haitaowang,liangcj,zhanghaipeng,hujiawei,sichengwang\}@stu.xmu.edu.cn}, \\
    \texttt{\{siqishen,clwen,cwang\}@xmu.edu.cn},
  \end{tabular}
}

\begin{document}

\maketitle

\begin{abstract}
Automated multi-agent systems offer clear advantages over manually designed ones in scalability and adaptability, but existing workflow topology methods still face important limitations. Search-based methods are often computationally expensive, textual-gradient-based methods rely on coarse-grained feedback, and existing generation-based methods are not well suited to discrete workflow topologies with complex dependencies. To address these limitations, we propose FlowMAS, a multi-agent workflow topology method based on Generative Flow Networks (GFlowNets). FlowMAS models workflow generation as reward-guided flow over the topology space and introduces three components: a GFlowNet-based topology generation backbone, a curiosity-driven module for structure-aware exploration, and an information-guided optimization module for evaluating intermediate topologies. Concretely, the curiosity-driven module encourages exploration of structurally novel workflows, while the information-guided module measures both the information contribution and the communication efficiency of different operators to favor more informative and effective collaboration patterns. Experiments on six benchmark datasets with three LLM backbones show that FlowMAS consistently outperforms multiple baselines. 

\end{abstract}

\input{paper_sections/introduction.tex}
\input{paper_sections/background.tex}
\input{paper_sections/method.tex}
\input{paper_sections/experiment.tex}

\input{paper_sections/related_work.tex}

\section{Conclusion}
In this work, we propose FlowMAS, which combines a generative flow network topology generation backbone, an information-guided optimization module for evaluating intermediate topologies, and a curiosity-driven module for structure-aware exploration. Together, these components address both the structural challenge of exploring diverse workflow topologies and the information challenge of identifying effective collaboration patterns among operators. Extensive experiments show that FlowMAS can generate more diverse and task-adaptive workflows and consistently outperforms strong baselines across different LLM backbones.

\section{Acknowledgments}
This work was partially supported by a grant from the National Key Laboratory of Unmanned Aerial Vehicle Technology Intelligence Development Fund. We thank the anonymous reviewer for their valuable suggestions.

\bibliographystyle{unsrt}
{\small
\bibliography{references}
}


\appendix




\newpage
\input{paper_sections/appendix.tex}

\newpage
\input{paper_sections/checklist.tex}

\end{document}

%% file: paper_sections/introduction.tex
\section{Introduction}
As the core carrier for coordinating multi-agent collaboration and decomposing complex tasks, agentic workflows have been widely applied to complex scenarios such as multi-hop question answering~\cite{hotpotqa_benchmark, drop_benchmark, maas}, code generation~\cite{humaneval_benchmark, mbpp_benchmark,metagpt}, and mathematical reasoning~\cite{gsm8k_benchmark, math_benchmark, llm_majority}. Agentic workflow topology~\cite{workflowsurvey} refers to how multiple agents or operators are organized and connected to coordinate reasoning, communication, and decision making toward a task. Its design~\cite{maas,gdesigner} largely determines the quality of collaboration, the efficiency of execution, and the overall capability of the system. 

Existing agentic workflow topology methods can be categorized into (1) manually crafted methods such as MetaGPT~\cite{metagpt}, LLM-Majority~\cite{llm_majority}, and LLM-Debate~\cite{llm_debate}; (2) search-based methods that search the workflow topology in a search space, such as ADAS~\cite{adas} and AFlow~\cite{aflow}; (3) textual-gradient-based methods that optimize workflow topology through an LLM-based critic/rectifier, such as OneFlow~\cite{oneflow} and DebFlow~\cite{DebFlow}; and (4) generation-based methods that use a variational autoencoder (VAE)~\cite{gdesigner}, graph diffusion~\cite{gtd}, or a multi-layer perceptron (MLP)~\cite{maas} to generate workflows. Generation-based methods are the best-performing genre of topology methods~\cite{maas, gtd}.

\begin{figure}[!t]
\centering
    \includegraphics[width=\linewidth]{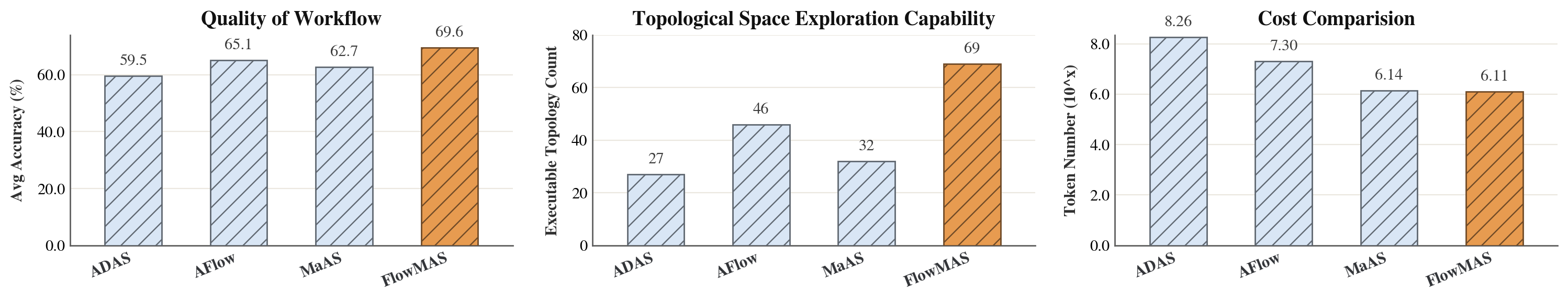}
    \captionsetup{hypcap=false}
    \captionof{figure}{Experiments with Llama-3.1-8B-Instruct on HumanEval. \textbf{Left:} the quality of the generated workflow for solving specific test problems. \textbf{Middle:} the number of executable workflows discovered during topology exploration on the training set. \textbf{Right:} Total token consumption.}
    \label{fig:intro_cmp}
\end{figure}

Manually crafted methods are labor-intensive~\cite{metagpt} and costly to optimize for a specific query. Search-based methods need to explore a large design space (e.g., a search tree~\cite{aflow}), which could be token-expensive. As the number of agentic operators increases, the performance of search-based methods could drop while token usage increases significantly. Textual gradient-based methods (e.g., OneFlow~\cite{oneflow}) are suboptimal because they usually optimize workflows through coarse-grained textual feedback, making it difficult to assign credit to specific workflow structures. Generation-based methods use generative backbones, such as VAEs~\cite{VAE} and diffusion models~\cite{gtd}, to construct workflows. When trained on curated workflow datasets, these methods may inherit the structural biases of the training data, limiting the diversity of the generated topologies. Moreover, backbones originally developed for image~\cite{stablediffusion} or vector generation~\cite{DDPM} may not adequately capture the discrete structures and complex dependencies of agentic workflows. Together, these limitations can restrict the diversity of workflows produced by existing search-based and generation-based methods, hindering their adaptation to task-specific requirements, as shown in Figure~\ref{fig:intro_cmp}.

To address the above workflow topology generation challenges, we propose FlowMAS, a multi-agent workflow topology method which adopts Generative Flow Networks (GFlowNets)~\cite{gflownet,bengio2021flow,malkin2022trajectory} as the topology generation backbone. GFlowNets are a generative modeling framework whose core idea is to model the generation process as a trajectory flow on a Directed Acyclic Graph (DAG). Thanks to their efficient generation capability in discrete combinatorial spaces and reward-driven sampling characteristics, GFlowNets are a promising solution for overcoming the existing limitations.


Despite the above advantages, directly applying GFlowNets to agentic workflows faces two challenges: an information challenge and a structure challenge. Information from multiple agentic operators could be unique, redundant, or synergistic. As a method developed for molecular generation, GFlowNets fail to model the information exchange among agentic operators. However, the information exchange in workflow topology is key to task success~\cite{workflowsurvey}. Regarding the structure challenge, GFlowNets could be inefficient in finding diverse topology structure (shown in Figure~\ref{fig:intro_cmp} and Tab.~\ref{tab:ablation_pid}) which can lead to suboptimial task performance.


FlowMAS consists of three key modules: a GFlowNet-based topology generation backbone, an information-guided optimization module for evaluating intermediate topologies and a curiosity-driven module for structure-aware exploration. Specifically, the GFlowNet backbone models workflow generation as a reward-guided flow over the topology space, the curiosity-driven module encourages exploration of structurally novel workflows during training, and the information-guided module measures the information contribution and communication efficiency of different operators to favor more informative and effective collaboration patterns.

Through extensive experiments on multiple benchmarks with multiple LLM backbones, we show that FlowMAS achieves the best performance among 15 baselines, thanks to its ability to generate diverse and query-specific workflows. Through ablation studies, we show that the partial information decomposition function could lead to better information sharing and that the curiosity-driven metric could lead to more diverse workflow topologies.


%% file: paper_sections/background.tex
\section{Background}

\subsection{Generative Flow Networks}
Generative Flow Networks (GFlowNets)~\cite{gflownet} are a probabilistic framework for training stochastic policies to generate discrete, combinatorial objects, originally developed to explore the diversity of molecular combinations in drug discovery. The generation process is modeled over a directed acyclic graph (DAG) $\mathcal{G} = (\mathcal{S}, \mathcal{A})$, where $\mathcal{S}$ denotes the state space and $\mathcal{A}$ denotes the set of valid state transitions. The goal of a GFlowNet is to learn a forward transition policy $P_F(s_{t+1}|s_t)$ such that the marginal probability of any terminal state $s_n$ is proportional to the reward function $R(x)$. A backward policy $P_B(s_t|s_{t+1})$ can also be defined to model the reverse process. Each edge is associated with a flow $F(s \rightarrow s')$, and the flow is \emph{consistent} if, for all internal states $s$, the incoming flow equals the outgoing flow:

\begin{equation}
\sum_{s'' \rightarrow s} F(s'' \rightarrow s) = F(s) = \sum_{s \rightarrow s'} F(s \rightarrow s').
\end{equation}

GFlowNet methods offer several advantages when generating complex discrete structures. GFlowNets can generate high-quality discrete structures from scratch using only a reward function $R(\cdot)$, naturally supporting graphs and workflow substructures. Their explicit modeling of the construction process as sequences of decisions over a DAG allows consideration of dependencies and logical constraints, making them particularly well-suited for automatic exploration and generation of multi-agent workflows.

\subsection{Partial Information Decomposition}

Partial Information Decomposition (PID)~\cite{williams2010nonnegative} decomposes the information that multiple sources $\{X_k\}_{k=1}^{K}$ carry about a target $Y$ into unique, redundant, and synergistic components. For two sources, the decomposition can be written as
\begin{equation}
I(\{X_i,X_j\};Y)
=
UI_i + UI_j + Red_{ij} + Syn_{ij},
\end{equation}
where $UI$, $Red$, and $Syn$ denote unique, shared, and synergistic target-relevant information, respectively. For multiple sources, PID further considers higher-order redundancy and synergy among source subsets. This framework allows quantifying the unique, shared, and synergistic contributions of multiple sources to a target, providing a principled tool to analyze redundancy and synergy in multi-agent systems or complex workflows.

%% file: paper_sections/method.tex
\begin{figure}[!t]
  \centering
  \includegraphics[width=1\textwidth]{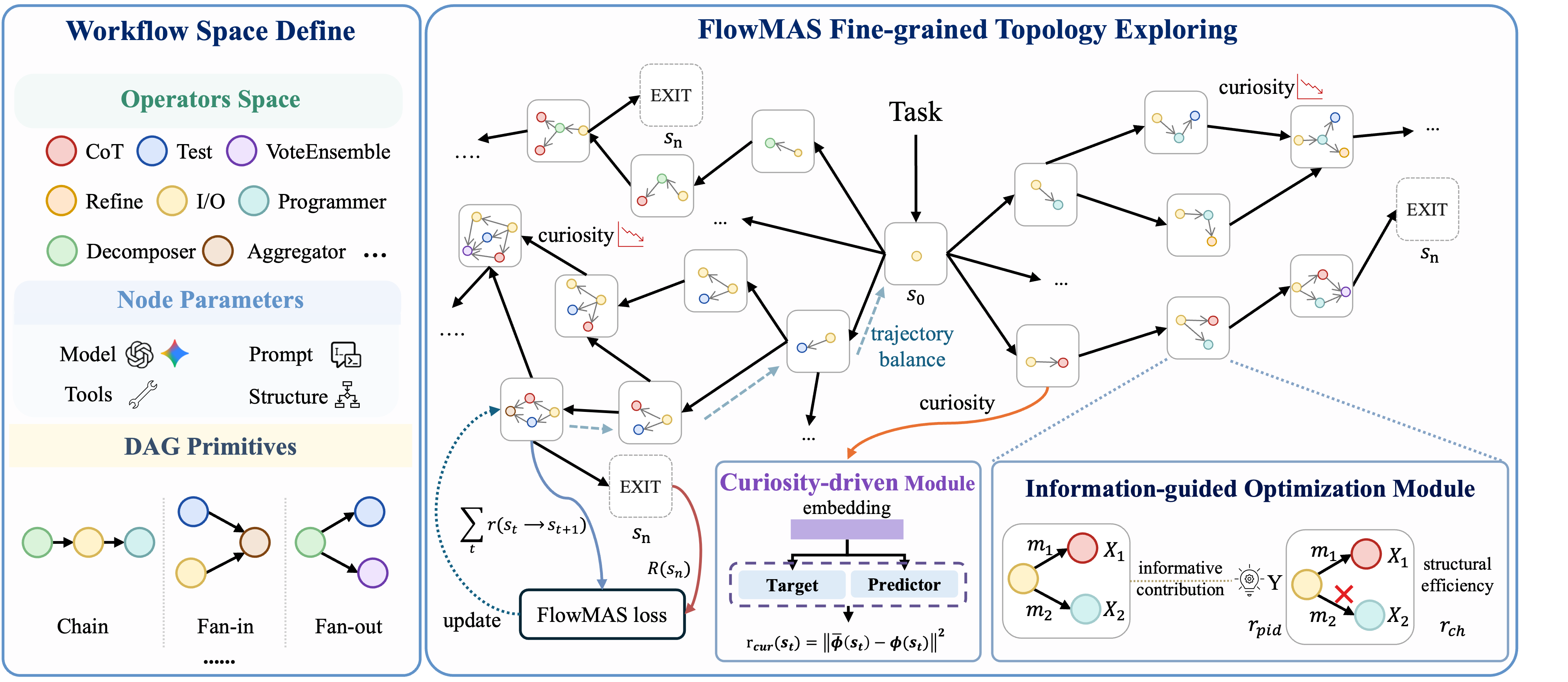}
  \caption{The overall framework of our proposed FlowMAS.}
  \label{fig:method}
\end{figure}

\section{Methodology}
Figure~\ref{fig:method} shows the overall framework of FlowMAS. Given a task input, FlowMAS uses GFlowNets to model the workflow distribution, a curiosity-driven module(CDM) to encourage structural exploration, and an information-guided optimization module(IOM) to favor high-quality workflows. The rest of this section presents the problem formulation in Section~\ref{sec:problem_formulation}, the FlowMAS exploration mechanism in Section~\ref{sec:3.2}, and the information-guided optimization strategy in Section~\ref{sec:3.3}.

\subsection{Problem Formulation}
\label{sec:problem_formulation}
\paragraph{Agentic Workflow.} We formulate the automated construction of LLM-based agentic systems as a workflow topology generation problem. Specifically, a workflow is represented as a directed acyclic graph (DAG) $\mathcal{W}=(\mathcal{V},\mathcal{E})$, where $\mathcal{V}=\{v_{1},v_{2},...,v_{n}\}$ denotes the set of nodes and $\mathcal{E}$ denotes the set of directed edges. Each node $v_{i}$ corresponds to an operator, i.e., a basic computational unit in the workflow. An operator is specified by its model $\mathcal{M}$, prompt $\mathcal{P}$, tool usage $\mathcal{T}$, and computational structure $\mathcal{C}$, where the computational structure defines how the operator runs internally, such as \texttt{CoT}, \texttt{Reflexion}, or \texttt{Programmer}.

A directed edge $e_{ij}=(v_{i},v_{j})$ represents the information flow from node $v_{i}$ to node $v_{j}$. In practice, we use three basic topology primitives to describe agentic workflows: chain, fan-in, and fan-out. A chain connects nodes sequentially and forms a linear pipeline. A fan-in merges information from multiple upstream nodes into a single downstream node, while a fan-out sends one node's output to multiple downstream nodes for parallel exploration. More complex topologies, such as tree, diamond, and hierarchical structures, can be constructed by combining these primitives.

\paragraph{Automated Workflow Topology Generation.} Given a query $q$ and an evaluation function $\mathcal{R(\cdot)}$ that measures the quality of a workflow $\mathcal{W}$ generated for query $q$ from the perspectives of effectiveness and efficiency, the goal of the proposed method is to explore the combinatorial state space of workflows represented as DAG structures ($\mathcal{W}=(\mathcal{V},\mathcal{E})$) and to identity high-quality solutions, as shown below:

\begin{equation}
\mathcal{P}_\theta(\mathcal{W}) \propto \mathcal{R(\mathcal{W})},
\label{equ:4}
\end{equation}


where $\mathcal{P_\theta(\cdot)}$ is the topology generation strategy. 

\subsection{FlowMAS}
\label{sec:3.2}

FlowMAS learns a flow distribution function $Z_{\theta}$ such that the total flow of all generation paths from the source state $s_{0}$ to a terminal state $s_{n}$ satisfies Equation~\ref{equ:4}, i.e., the flow is proportional to the final value (e.g., task accuracy, token cost). Consequently, GFlowNets allocates limited flow across different structural paths, allowing high-value structures to receive higher flow while maintaining exploration diversity.

In FlowMAS, the workflow construction process is denoted as a trajectory $\tau=\{s_{0},s_{1},...,s_{n}\}$, where each state $s_{t}$ represents a partially built workflow topology represented by DAG. Each state $s_{t}$ is uniformly encoded as:

\begin{equation}
h(s_{t}) = Emb(e_{task},e_{op},e_{topo}),
\end{equation}

where $Emb(\cdot)$ is an embedding model, such as MiniLM~\cite{minilm}. The terms $e_{task}$, $e_{op}$, $e_{topo}$ denote the embeddings of the current task text, the operator pool, and the composed workflow topology, respectively. Based on the current state, the model performs actions $a_{t}$ to add nodes and connect edges, thereby gradually expanding the information flow paths.

However, directly applying GFlowNets to agentic workflow generation does not fully address the two challenges highlighted in the introduction. For the \emph{structural challenge}, relying only on the terminal reward $r(s_{n})$ provides little guidance about whether an intermediate topology is worth expanding, which can lead to redundant structures and insufficient exploration of diverse topologies. For the \emph{information challenge}, the terminal reward does not indicate whether the current partial workflow already captures useful coordination patterns among operators, such as complementary or synergistic information exchange. As a result, promising intermediate structures may receive too little flow during generation.

To address these issues, the core idea of FlowMAS is to introduce \emph{intermediate flow} into the generation process. Instead of assigning value only to the final workflow, we assign reward signals to intermediate transitions so that the flow can reflect both structural quality and information quality throughout topology construction. In this way, the model is encouraged to allocate more probability mass to trajectories whose intermediate states are already structurally efficient and information-effective, rather than waiting until the final reward to distinguish good workflows from bad ones.

Formally, we introduce an intermediate reward $r(s_{t} \rightarrow s_{t+1})$ on each transition and treat it as reward-induced local flow injected into the trajectory. This additional flow reshapes the original flow conservation constraint as follows:

\begin{equation}
 \sum_{s_{t-1}} F(s_{t-1} \rightarrow s_{t})=F(s_{t})=\sum_{s_{t+1}}[F(s_{t}\rightarrow s_{t+1})+r(s_{t} \rightarrow s_{t+1})], 
\label{equ:flow_match_cons}
\end{equation}

here, $F(s_{t})$ is the state flow of state $s_{t}$, and $F(s_{t} \rightarrow s_{t+1})$ is the edge flow for the transition from $s_{t}$ to $s_{t+1}$. Correspondingly, the trajectory balance constraint can be expressed as:

\begin{equation}
  Z_{\theta} \prod_{t=0}^{n-1} P_F(s_{t+1}|s_t) = R(s_{n}) \prod_{t=0}^{n-1} \left[ P_B(s_t|s_{t+1}) + \frac{r(s_t \to s_{t+1})}{F(s_{t+1})} \right], 
\label{equ:intermediate_reward}
\end{equation}

here, $P_{F}(\cdot)$ and $P_{B}(\cdot)$ denote the forward policy and the backward policy, respectively. A detailed derivation is provided in Appendix~\ref{app:TBC}. 

\subsection{Information-guided Topology Optimization}
\label{sec:3.3}

In this section, we describe the information-guided optimization module and curiosity-driven module in FlowMAS and specify the components of the intermediate reward $r(s_{t} \rightarrow s_{t+1})$ used to guide topology generation. 

\paragraph{Information-guided Optimization Module (IOM)} 
To address the information challenge, FlowMAS evaluates whether the local collaboration patterns formed during topology construction provide target-relevant and non-redundant information.

For each training instance, let $Y$ denote the representation of its ground-truth target answer. 
Given a local DAG primitive $D_i$ consisting of $K$ upstream operators whose messages are jointly consumed by a downstream operator, we introduce a PID-inspired synergistic information reward:
\begin{equation}
r_{pid}(D_i)
=
\max{(0, 
I(D_i;Y)
-
\sum_{k=1}^{K} UI_k(D_i;Y)
-
Red(D_i;Y))
} \in {[0,1]},
\label{eq:pid_reward}
\end{equation}
where $I(D_i,Y)$ measures the target alignment of the aggregated incoming messages, $UI_k(D_i;Y)$ represents the target-specific contribution unique to the $k$-th message, and $Red(D_i;Y)$ captures the target-aligned information shared across incoming messages. 
Intuitively, $r_{pid}(D_i)$ favors local collaboration patterns whose aggregated messages exhibit stronger target relevance than can be explained by isolated message contributions and their shared component alone. 
The detailed computation of these terms is provided in Appendix~\ref{app:pid_reward}.

Besides synergistic information, we also evaluate whether a topology uses its communication paths efficiently. To capture this property, we introduce an effective information channel reward. For the $K$ upstream operator outputs in $D_i$, we construct a cosine-similarity Gram matrix $\mathcal{M}$ and compute its spectral entropy from the eigenvalues $\lambda$ to obtain

\begin{equation}
r_{ch}(D_i)
=
\frac{2^{H(\rho)}-1}{K-1},
\quad
\text{where}
\quad
H(\rho)
=
-\sum_{j=1}^{K} \lambda_j \log_2 \lambda_j,
\end{equation}

where $r_{ch}(\cdot)$ is the effective channel reward, $H(\cdot)$ is the spectral entropy, and $\rho$ is the trace-normalized Gram matrix (see Appendix~\ref{app:SE} for details). Intuitively, this reward measures how evenly information is distributed across the communication channels of a substructure. A low reward indicates that information is concentrated in a few dominant paths, suggesting redundancy or inactive branches. A high reward indicates that multiple paths contribute more evenly, implying better coordination efficiency and less wasted computation. In multi-agent workflows, this term discourages redundant agents or edges and favors structures in which different agents make more balanced and effective contributions.

\paragraph{Curiosity-driven Module (CDM)} 
To address the structural challenge, we introduce a curiosity-driven module that encourages exploration of under-visited topologies. It assigns higher reward to structurally novel states, improving the coverage of potentially useful workflow structures during training. Specifically, the structural reward is defined as
$r_{\text{cur}}(s_t)=\| \bar{\phi}(s_{t}) - \phi(s_{t}) \|^2$,
where $\phi(\cdot)$ is a fixed target network and $\bar{\phi}(\cdot)$ is a trainable prediction network.

We combine the curiosity-driven module and the information-guided optimization module into a unified transition reward for intermediate topology evaluation. Let $\mathcal{D}_t=\{\mathcal{D}_{t,1},...,\mathcal{D}_{t,n}\}$ denote the set of local DAG primitives evaluated at the transition $s_t \rightarrow s_{t+1}$:

\begin{equation}
r(s_{t} \rightarrow s_{t+1})
=
\alpha r_{\text{cur}}(s_{t})
+
\beta
\sum_{D_i \in \mathcal{D}_t}
\left(
r_{pid}(D_i) + r_{ch}(D_i)
\right),
\label{equ:reward}
\end{equation}

where $\alpha$ and $\beta$ balance the structure-aware exploration reward and the information-guided optimization reward, respectively.

%% file: paper_sections/experiment.tex
\section{Experiment}
\subsection{Experimental Setup}

\paragraph{Tasks and Benchmarks.}
We evaluate \method{} on six public benchmarks in our main experiments. Following AFlow~\cite{aflow}, we use the same data partition protocol, splitting each benchmark into validation and test sets at a 1:4 ratio. For mathematical reasoning, we use GSM8K~\cite{gsm8k_benchmark} and MATH~\cite{math_benchmark}. For MATH, we select 605 level-5 problems from four representative categories: counting and probability, number theory, prealgebra, and precalculus. For code generation, we use HumanEval~\cite{humaneval_benchmark} and MBPP~\cite{mbpp_benchmark}. For reading comprehension, we use HotpotQA~\cite{hotpotqa_benchmark} and DROP~\cite{drop_benchmark}, from which we randomly sample 1{,}000 examples each, following Hu et al.~\cite{adas}. Detailed dataset statistics are provided in Appendix~\ref{app:benchmark_statistics}. In addition, we evaluate the cross-task generalization ability of \method{} on BIG-Bench Hard (BBH)~\cite{bbh_benchmark}, with results presented in Section~\ref{sec:generalization_analysis}.

\paragraph{Baselines.}
We compare \method{} with three series of agentic baselines: (1) single-agent test-time scaling methods, including Vanilla LLM (IO), CoT~\cite{cot}, CoT-SC~\cite{sc}, ReAct~\cite{react}, and Reflexion~\cite{reflexion}; (2) hand-crafted multi-agent systems, including LLM-Majority~\cite{llm_majority}, LLM-Debate~\cite{llm_debate}, LLM-Blender~\cite{llm_blender}, DyLAN~\cite{dylan}, and AgentVerse~\cite{agentverse}; (3) automated multi-agent systems, including ADAS~\cite{adas}, AgentNet~\cite{agentnet}, SELFORG~\cite{selforg}, AFlow~\cite{aflow}, and MaAS~\cite{maas}. More details on baseline setups are provided in Appendix~\ref{app:baseline_methods_detail}.

\paragraph{LLM backbone.}
 All methods are evaluated on three backbone LLMs, spanning one closed-source LLM (gpt-4o-mini~\cite{openai2024gpt4o}) and two open-source LLMs (Llama-3.1-8B-Instruct~\cite{grattafiori2024llama}, Gemma-4-26B-A4B-it~\cite{google2025gemma4}).
 For all backbone LLMs, we set the sampling temperature to 0.2. 

\paragraph{Metrics.}
For GSM8K and MATH, we report \emph{solve rate} (\%). For HumanEval and MBPP, we report \emph{pass@1}. For HotpotQA and DROP, we report \emph{F1}. For all methods, we additionally track total token usage as the cost metric. 
For each experiment type, we run five independent trials and report the mean and standard deviation.

\subsection{Performance Analysis}
Table~\ref{tab:main_llama} compares \method{} with 15 baselines on six benchmarks. Table~\ref{tab:main_other_backbones} further reports the results of \method{}, IO, and five representative automated workflow-design baselines on GPT-4o-mini and Gemma-4-26B-A4B-it; the full comparisons with all baselines on these two backbones are provided in Appendix~\ref{app:additional_backbone_results}. Averaging over all methods in each category and all three backbones, \method{} surpasses the hand-crafted multi-agent methods by 8.17 points and the existing automated workflow-design methods by 6.12 points. On Llama-3.1-8B-Instruct, \method{} outperforms AgentNet and SELFORG by 11.07 points on average, suggesting that its performance is less dependent on strong backbone capabilities. Moreover, compared with MaAS and AFlow, \method{} maintains consistent advantages across all three backbones, demonstrating stronger generality and overall effectiveness.

\begin{table*}[!t]
    \centering
    \renewcommand{\arraystretch}{1.12}
    \caption{Main results on Llama-3.1-8B-Instruct. We \best{bold} the best results and \second{underline} the second-best results. Each experiment is run five times to record the mean and standard deviation.}
    \label{tab:main_llama}
    \resizebox{\textwidth}{!}{%
    \begin{tabular}{l|ccccccc}
        \toprule
        \rowcolor{black!4}
        Method & HumanEval & GSM8K & MATH & MBPP & HotpotQA & DROP & Avg. \\
        \midrule
        \rowcolor{black!4}
        IO & 50.1\textcolor{SlateBlue}{\scriptsize$\pm 1.4$} & 73.7\textcolor{SlateBlue}{\scriptsize$\pm 1.4$} & 18.9\textcolor{SlateBlue}{\scriptsize$\pm 1.5$} & 67.1\textcolor{SlateBlue}{\scriptsize$\pm 1.3$} & 60.5\textcolor{SlateBlue}{\scriptsize$\pm 1.6$} & 68.8\textcolor{SlateBlue}{\scriptsize$\pm 2.1$} & 56.52 \\
        CoT~\cite{cot} & 51.2\textcolor{SlateBlue}{\scriptsize$\pm 1.5$} & 75.6\textcolor{SlateBlue}{\scriptsize$\pm 1.1$} & 21.0\textcolor{SlateBlue}{\scriptsize$\pm 1.7$} & 69.4\textcolor{SlateBlue}{\scriptsize$\pm 1.4$} & 64.0\textcolor{SlateBlue}{\scriptsize$\pm 1.9$} & 73.1\textcolor{SlateBlue}{\scriptsize$\pm 1.9$} & 59.05 \\
        \rowcolor{black!4}
        CoT-SC~\cite{sc} & 52.4\textcolor{SlateBlue}{\scriptsize$\pm 0.8$} & 77.5\textcolor{SlateBlue}{\scriptsize$\pm 0.9$} & 22.1\textcolor{SlateBlue}{\scriptsize$\pm 0.0$} & 66.8\textcolor{SlateBlue}{\scriptsize$\pm 0.8$} & 66.0\textcolor{SlateBlue}{\scriptsize$\pm 0.9$} & 75.4\textcolor{SlateBlue}{\scriptsize$\pm 1.3$} & 60.03 \\
        ReAct~\cite{react} & 51.9\textcolor{SlateBlue}{\scriptsize$\pm 1.9$} & 75.2\textcolor{SlateBlue}{\scriptsize$\pm 1.7$} & 28.1\textcolor{SlateBlue}{\scriptsize$\pm 2.0$} & 65.4\textcolor{SlateBlue}{\scriptsize$\pm 1.5$} & 63.7\textcolor{SlateBlue}{\scriptsize$\pm 1.7$} & 71.9\textcolor{SlateBlue}{\scriptsize$\pm 1.7$} & 59.37 \\
        \rowcolor{black!4}
        Reflexion~\cite{reflexion} & 50.5\textcolor{SlateBlue}{\scriptsize$\pm 1.7$} & 79.1\textcolor{SlateBlue}{\scriptsize$\pm 1.1$} & 22.0\textcolor{SlateBlue}{\scriptsize$\pm 2.1$} & 65.7\textcolor{SlateBlue}{\scriptsize$\pm 1.4$} & 57.6\textcolor{SlateBlue}{\scriptsize$\pm 1.5$} & 64.6\textcolor{SlateBlue}{\scriptsize$\pm 2.0$} & 56.58 \\
        \groupsep
        LLM-Majority~\cite{llm_majority} & 56.6\textcolor{SlateBlue}{\scriptsize$\pm 0.0$} & 76.7\textcolor{SlateBlue}{\scriptsize$\pm 1.2$} & 22.4\textcolor{SlateBlue}{\scriptsize$\pm 1.6$} & 67.3\textcolor{SlateBlue}{\scriptsize$\pm 0.9$} & 62.0\textcolor{SlateBlue}{\scriptsize$\pm 1.1$} & 71.4\textcolor{SlateBlue}{\scriptsize$\pm 1.2$} & 59.40 \\
        \rowcolor{black!4}
        LLM-Debate~\cite{llm_debate} & 53.4\textcolor{SlateBlue}{\scriptsize$\pm 1.3$} & 82.7\textcolor{SlateBlue}{\scriptsize$\pm 1.4$} & 19.9\textcolor{SlateBlue}{\scriptsize$\pm 1.7$} & 65.4\textcolor{SlateBlue}{\scriptsize$\pm 1.2$} & 66.9\textcolor{SlateBlue}{\scriptsize$\pm 1.6$} & 64.4\textcolor{SlateBlue}{\scriptsize$\pm 1.4$} & 58.78 \\
        LLM-Blender~\cite{llm_blender} & 52.5\textcolor{SlateBlue}{\scriptsize$\pm 1.5$} & 82.1\textcolor{SlateBlue}{\scriptsize$\pm 1.9$} & 22.3\textcolor{SlateBlue}{\scriptsize$\pm 1.8$} & 64.5\textcolor{SlateBlue}{\scriptsize$\pm 1.4$} & 68.0\textcolor{SlateBlue}{\scriptsize$\pm 1.8$} & 75.6\textcolor{SlateBlue}{\scriptsize$\pm 1.6$} & 60.83 \\
        \rowcolor{black!4}
        DyLAN~\cite{dylan} & 51.9\textcolor{SlateBlue}{\scriptsize$\pm 1.4$} & 83.0\textcolor{SlateBlue}{\scriptsize$\pm 1.7$} & 27.6\textcolor{SlateBlue}{\scriptsize$\pm 2.0$} & 68.3\textcolor{SlateBlue}{\scriptsize$\pm 1.5$} & 63.2\textcolor{SlateBlue}{\scriptsize$\pm 1.9$} & 70.2\textcolor{SlateBlue}{\scriptsize$\pm 1.8$} & 60.70 \\
        AgentVerse~\cite{agentverse} & 52.4\textcolor{SlateBlue}{\scriptsize$\pm 1.7$} & 76.9\textcolor{SlateBlue}{\scriptsize$\pm 2.1$} & 26.7\textcolor{SlateBlue}{\scriptsize$\pm 2.1$} & 66.3\textcolor{SlateBlue}{\scriptsize$\pm 1.6$} & 68.2\textcolor{SlateBlue}{\scriptsize$\pm 1.5$} & 70.8\textcolor{SlateBlue}{\scriptsize$\pm 1.6$} & 60.22 \\
        \groupsep
        \rowcolor{black!4}
        ADAS~\cite{adas} & 59.5\textcolor{SlateBlue}{\scriptsize$\pm 2.4$} & 79.5\textcolor{SlateBlue}{\scriptsize$\pm 1.8$} & 22.0\textcolor{SlateBlue}{\scriptsize$\pm 3.0$} & 64.0\textcolor{SlateBlue}{\scriptsize$\pm 1.4$} & 60.9\textcolor{SlateBlue}{\scriptsize$\pm 2.9$} & 69.0\textcolor{SlateBlue}{\scriptsize$\pm 2.6$} & 59.15 \\
        AgentNet~\cite{agentnet} & 50.8\textcolor{SlateBlue}{\scriptsize$\pm 2.2$} & 75.0\textcolor{SlateBlue}{\scriptsize$\pm 0.4$} & 22.5\textcolor{SlateBlue}{\scriptsize$\pm 0.1$} & 60.4\textcolor{SlateBlue}{\scriptsize$\pm 0.4$} & 61.7\textcolor{SlateBlue}{\scriptsize$\pm 0.1$} & 67.9\textcolor{SlateBlue}{\scriptsize$\pm 0.2$} & 56.38 \\
        \rowcolor{black!4}
        SELFORG~\cite{selforg} & 61.9\textcolor{SlateBlue}{\scriptsize$\pm 0.6$} & 83.1\textcolor{SlateBlue}{\scriptsize$\pm 0.1$} & 26.0\textcolor{SlateBlue}{\scriptsize$\pm 0.2$} & 67.0\textcolor{SlateBlue}{\scriptsize$\pm 0.5$} & 65.7\textcolor{SlateBlue}{\scriptsize$\pm 0.1$} & 71.2\textcolor{SlateBlue}{\scriptsize$\pm 0.1$} & 62.48 \\
        AFlow~\cite{aflow} & \second{65.1\textcolor{SlateBlue}{\scriptsize$\pm 1.4$}} & \second{85.1\textcolor{SlateBlue}{\scriptsize$\pm 1.6$}} & 35.3\textcolor{SlateBlue}{\scriptsize$\pm 2.2$} & 72.9\textcolor{SlateBlue}{\scriptsize$\pm 1.4$} & 69.2\textcolor{SlateBlue}{\scriptsize$\pm 1.9$} & \second{77.7\textcolor{SlateBlue}{\scriptsize$\pm 2.1$}} & 67.55 \\
        \rowcolor{black!4}
        MaAS~\cite{maas} & 62.7\textcolor{SlateBlue}{\scriptsize$\pm 0.9$} & 85.0\textcolor{SlateBlue}{\scriptsize$\pm 1.1$} & \second{37.5\textcolor{SlateBlue}{\scriptsize$\pm 1.9$}} & \second{73.4\textcolor{SlateBlue}{\scriptsize$\pm 1.0$}} & \second{70.3\textcolor{SlateBlue}{\scriptsize$\pm 1.6$}} & 77.5\textcolor{SlateBlue}{\scriptsize$\pm 1.2$} & \second{67.73} \\
        \groupsep
        \rowcolor{methodhighlight}
        \method{} & \best{69.6\textcolor{SlateBlue}{\scriptsize$\pm 1.5$}} & \best{86.2\textcolor{SlateBlue}{\scriptsize$\pm 1.0$}} & \best{38.8\textcolor{SlateBlue}{\scriptsize$\pm 1.8$}} & \best{75.6\textcolor{SlateBlue}{\scriptsize$\pm 1.1$}} & \best{73.6\textcolor{SlateBlue}{\scriptsize$\pm 1.2$}} & \best{79.2\textcolor{SlateBlue}{\scriptsize$\pm 1.4$}} & \best{70.50} \\
        \bottomrule
    \end{tabular}%
    }

\end{table*}

\begin{table*}[!t]
    \centering
    \renewcommand{\arraystretch}{1.12}
    \caption{Results of IO and selected automated MAS-design methods on Gemma-4-26B-A4B-it and GPT-4o-mini backbones. We \best{bold} the best results and \second{underline} the second-best results.}
    \label{tab:main_other_backbones}
    \resizebox{\textwidth}{!}{%
    \begin{tabular}{l|ccccccc}
        \toprule
        \rowcolor{black!4}
        Method & HumanEval & GSM8K & MATH & MBPP & HotpotQA & DROP & Avg. \\
        \midrule
        \rowcolor{black!8}
        \multicolumn{8}{c}{\textbf{GPT-4o-mini}} \\
        
        \rowcolor{black!4}
        IO & 86.3\textcolor{SlateBlue}{\scriptsize$\pm 0.6$} & 86.3\textcolor{SlateBlue}{\scriptsize$\pm 0.5$} & 45.0\textcolor{SlateBlue}{\scriptsize$\pm 0.7$} & 71.2\textcolor{SlateBlue}{\scriptsize$\pm 0.4$} & 66.7\textcolor{SlateBlue}{\scriptsize$\pm 0.6$} & 73.2\textcolor{SlateBlue}{\scriptsize$\pm 0.9$} & 71.45 \\
        ADAS~\cite{adas} & 83.4\textcolor{SlateBlue}{\scriptsize$\pm 1.2$} & 84.9\textcolor{SlateBlue}{\scriptsize$\pm 0.9$} & 42.7\textcolor{SlateBlue}{\scriptsize$\pm 1.2$} & 67.3\textcolor{SlateBlue}{\scriptsize$\pm 0.8$} & 63.9\textcolor{SlateBlue}{\scriptsize$\pm 0.9$} & 75.5\textcolor{SlateBlue}{\scriptsize$\pm 1.2$} & 69.62 \\
        \rowcolor{black!4}
        AgentNet~\cite{agentnet} & 86.8\textcolor{SlateBlue}{\scriptsize$\pm 0.9$} & 91.8\textcolor{SlateBlue}{\scriptsize$\pm 0.0$} & 47.5\textcolor{SlateBlue}{\scriptsize$\pm 0.1$} & 71.1\textcolor{SlateBlue}{\scriptsize$\pm 0.2$} & 64.5\textcolor{SlateBlue}{\scriptsize$\pm 0.2$} & 80.8\textcolor{SlateBlue}{\scriptsize$\pm 0.0$} & 73.75 \\
        SELFORG~\cite{selforg} & 86.0\textcolor{SlateBlue}{\scriptsize$\pm 2.0$} & \second{92.7\textcolor{SlateBlue}{\scriptsize$\pm 0.1$}} & 50.7\textcolor{SlateBlue}{\scriptsize$\pm 0.7$} & 71.7\textcolor{SlateBlue}{\scriptsize$\pm 0.3$} & 68.9\textcolor{SlateBlue}{\scriptsize$\pm 0.0$} & \second{84.3\textcolor{SlateBlue}{\scriptsize$\pm 0.1$}} & 75.72 \\
        \rowcolor{black!4}
        AFlow~\cite{aflow} & 89.6\textcolor{SlateBlue}{\scriptsize$\pm 0.7$} & 90.5\textcolor{SlateBlue}{\scriptsize$\pm 0.7$} & 50.5\textcolor{SlateBlue}{\scriptsize$\pm 0.9$} & 81.0\textcolor{SlateBlue}{\scriptsize$\pm 0.6$} & 72.6\textcolor{SlateBlue}{\scriptsize$\pm 0.9$} & 79.7\textcolor{SlateBlue}{\scriptsize$\pm 0.5$} & 77.32 \\
        MaAS~\cite{maas} & \second{92.2\textcolor{SlateBlue}{\scriptsize$\pm 0.5$}} & 91.7\textcolor{SlateBlue}{\scriptsize$\pm 0.6$} & \second{51.1\textcolor{SlateBlue}{\scriptsize$\pm 0.8$}} & \second{81.4\textcolor{SlateBlue}{\scriptsize$\pm 0.5$}} & \second{77.3\textcolor{SlateBlue}{\scriptsize$\pm 0.7$}} & 83.1\textcolor{SlateBlue}{\scriptsize$\pm 0.6$} & \second{79.47} \\
        \rowcolor{methodhighlight}
        \method{} & \best{94.4\textcolor{SlateBlue}{\scriptsize$\pm 0.8$}} & \best{93.1\textcolor{SlateBlue}{\scriptsize$\pm 0.4$}} & \best{52.6\textcolor{SlateBlue}{\scriptsize$\pm 0.7$}} & \best{83.4\textcolor{SlateBlue}{\scriptsize$\pm 0.5$}} & \best{80.0\textcolor{SlateBlue}{\scriptsize$\pm 0.4$}} & \best{84.9\textcolor{SlateBlue}{\scriptsize$\pm 0.2$}} & \best{81.40} \\
        \midrule
        \rowcolor{black!8}
        \multicolumn{8}{c}{\textbf{Gemma-4-26B-A4B-it}} \\
        \rowcolor{black!4}
        IO & 90.8\textcolor{SlateBlue}{\scriptsize$\pm 0.8$} & 90.5\textcolor{SlateBlue}{\scriptsize$\pm 0.7$} & 61.3\textcolor{SlateBlue}{\scriptsize$\pm 0.9$} & 73.3\textcolor{SlateBlue}{\scriptsize$\pm 0.6$} & 74.9\textcolor{SlateBlue}{\scriptsize$\pm 0.8$} & 83.9\textcolor{SlateBlue}{\scriptsize$\pm 1.0$} & 79.12 \\
        ADAS~\cite{adas} & 88.5\textcolor{SlateBlue}{\scriptsize$\pm 1.2$} & 90.0\textcolor{SlateBlue}{\scriptsize$\pm 1.0$} & 59.2\textcolor{SlateBlue}{\scriptsize$\pm 1.2$} & 71.2\textcolor{SlateBlue}{\scriptsize$\pm 0.8$} & 76.7\textcolor{SlateBlue}{\scriptsize$\pm 1.1$} & 86.5\textcolor{SlateBlue}{\scriptsize$\pm 1.2$} & 78.68 \\
        \rowcolor{black!4}
        AgentNet~\cite{agentnet} & 90.9\textcolor{SlateBlue}{\scriptsize$\pm 1.2$} & 91.9\textcolor{SlateBlue}{\scriptsize$\pm 1.0$} & 62.7\textcolor{SlateBlue}{\scriptsize$\pm 0.3$} & 81.6\textcolor{SlateBlue}{\scriptsize$\pm 0.3$} & 79.5\textcolor{SlateBlue}{\scriptsize$\pm 0.4$} & 88.8\textcolor{SlateBlue}{\scriptsize$\pm 1.0$} & 82.57 \\
        SELFORG~\cite{selforg} & 91.3\textcolor{SlateBlue}{\scriptsize$\pm 0.3$} & 92.1\textcolor{SlateBlue}{\scriptsize$\pm 0.3$} & 63.1\textcolor{SlateBlue}{\scriptsize$\pm 0.4$} & 77.8\textcolor{SlateBlue}{\scriptsize$\pm 0.4$} & 79.5\textcolor{SlateBlue}{\scriptsize$\pm 0.5$} & 90.4\textcolor{SlateBlue}{\scriptsize$\pm 0.5$} & 82.37 \\
        \rowcolor{black!4}
        AFlow~\cite{aflow} & \second{92.6\textcolor{SlateBlue}{\scriptsize$\pm 0.8$}} & 93.4\textcolor{SlateBlue}{\scriptsize$\pm 0.9$} & 65.0\textcolor{SlateBlue}{\scriptsize$\pm 1.0$} & \second{83.5\textcolor{SlateBlue}{\scriptsize$\pm 0.7$}} & 82.7\textcolor{SlateBlue}{\scriptsize$\pm 1.0$} & 89.9\textcolor{SlateBlue}{\scriptsize$\pm 1.1$} & 84.52 \\
        MaAS~\cite{maas} & 92.4\textcolor{SlateBlue}{\scriptsize$\pm 0.6$} & \second{95.1\textcolor{SlateBlue}{\scriptsize$\pm 0.7$}} & \second{65.1\textcolor{SlateBlue}{\scriptsize$\pm 1.0$}} & 83.1\textcolor{SlateBlue}{\scriptsize$\pm 0.6$} & \second{83.8\textcolor{SlateBlue}{\scriptsize$\pm 0.8$}} & \second{90.7\textcolor{SlateBlue}{\scriptsize$\pm 0.7$}} & \second{85.03} \\
        \rowcolor{methodhighlight}
        \method{} & \best{94.8\textcolor{SlateBlue}{\scriptsize$\pm 0.8$}} & \best{95.6\textcolor{SlateBlue}{\scriptsize$\pm 0.8$}} & \best{66.7\textcolor{SlateBlue}{\scriptsize$\pm 0.9$}} & \best{87.1\textcolor{SlateBlue}{\scriptsize$\pm 0.6$}} & \best{85.3\textcolor{SlateBlue}{\scriptsize$\pm 0.8$}} & \best{92.0\textcolor{SlateBlue}{\scriptsize$\pm 0.7$}} & \best{86.92} \\
        \bottomrule
    \end{tabular}%
    }\end{table*}

\subsection{Cost Analysis}

To assess whether \method{} is cost-effective in both optimization and deployment, we report the training and inference token consumption of different baselines on HumanEval in Table~\ref{tab:humaneval_token_breakdown}. We focus on token usage and accuracy because they are the most direct and comparable measures of efficiency across methods.

\paragraph{Obs. 1: \method{}'s optimization is resource-friendly.} As shown in Table~\ref{tab:humaneval_token_breakdown}, among optimization-oriented agentic workflow methods, \method{} achieves the highest accuracy with relatively low training token consumption. MaAS uses fewer training tokens than \method{} but is 6.9 accuracy points lower. This accuracy--cost trade-off is consistent with the BBH analysis in Figure~\ref{fig:intro_cmp}, where MaAS explores substantially fewer executable topologies than \method{}.

\paragraph{Obs. 2: \method{} enjoys superior token economy during inference.} As shown in Table~\ref{tab:humaneval_token_breakdown}, AFlow uses more than 15$\times$ the inference tokens of \method{} and still fails to match its accuracy. AgentNet substantially reduces inference cost by reusing experience summarized during training, but the reused experience is less reliable because it is entirely induced by LLM summarization, leading to noticeably lower accuracy. In contrast, \method{} achieves the best overall performance with a relatively low token budget.

\subsection{Ablation Study}

We conduct two ablations to answer two questions. \textbf{Q1:} Is GFlowNet a better generation-based backbone for agentic flow than other generation models? \textbf{Q2:} Are the two key method components, the curiosity-driven module (CDM) and the information-guided optimization mechanism (IOM), introduced in Section~3.3 necessary, and can they not be replaced arbitrarily by other alternatives?
Tables~\ref{tab:ablation_backbone} and \ref{tab:ablation_pid} summarize the results.

\paragraph{Q1: Generator backbone ablation.} Table~\ref{tab:ablation_backbone} examines whether GFlowNet is a stronger core architecture than other advanced topology generators. To keep the comparison as clean as possible, we remove the curiosity-driven module (CDM) and information-guided optimization mechanism (IOM), and compare the GFlowNet backbone with a Graph Diffusion Model~\cite{DDPM, xu2024discrete}, a VAE~\cite{VAE}, a Transformer~\cite{transformer} and a MLP~\cite{mlp} controller under otherwise aligned settings, using GTD~\cite{gtd}, G-Designer~\cite{gdesigner}, and MaAS~\cite{maas} as references. Among the five generator-backbone variants, GFlowNet achieves the best result on all three benchmarks. This suggests that, relative to the alternative backbones, it provides a more suitable inductive bias for modeling complex discrete workflow structures with logical constraints. More specifically, the Graph Diffusion Model~\cite{DDPM} and the Transformer~\cite{transformer} obtain HumanEval accuracies below 60\%. The VAE attains 46.1 on HotpotQA, substantially below the 67.3 achieved by the GFlowNet variant.

\begin{table}[t]
    \centering
    \footnotesize
    \setlength{\tabcolsep}{4pt}
    \renewcommand{\arraystretch}{1.08}
    \caption{Efficiency comparison between \method{} and state-of-the-art baselines on the HumanEval. Train$\times$ and Infer$\times$ are relative to \method{}
    }
    \label{tab:humaneval_token_breakdown}
    \begin{tabular*}{\linewidth}{@{\extracolsep{\fill}}lrrrrr@{}}
        \toprule
        Method & Acc. (\%) & Training tokens & Train$\times$ & Inference tokens & Infer$\times$ \\
        \midrule
        DyLAN~\cite{dylan} & 51.9 & 5{,}842{,}507 & 15.2$\times$ & 25{,}064{,}472 & 27.8$\times$ \\
        ADAS~\cite{adas} & 59.5 & 84{,}526{,}998 & 220.0$\times$ & 97{,}526{,}460 & 108.3$\times$ \\
        AgentNet~\cite{agentnet} & 50.8 & 792{,}775 & 2.1$\times$ & 451{,}047 & 0.5$\times$ \\
        SELFORG~\cite{selforg} & 61.9 & 853{,}528 & 2.2$\times$ & 2{,}755{,}046 & 3.1$\times$ \\
        AFlow~\cite{aflow} & 65.1 & 6{,}253{,}861 & 16.3$\times$ & 13{,}690{,}205 & 15.2$\times$ \\
        MaAS~\cite{maas} & 62.7 & 223{,}042 & 0.6$\times$ & 1{,}151{,}392 & 1.3$\times$ \\
        \rowcolor{orange!8!white}
        \method{} (Ours) & 69.6 & 384{,}278 & 1.0$\times$ & 900{,}287 & 1.0$\times$ \\
        \bottomrule
    \end{tabular*}
\end{table}

\FloatBarrier

\begin{table}[H]
    \centering
    \small
    \setlength{\tabcolsep}{5pt}
    \renewcommand{\arraystretch}{1.02}
    \caption{Q1. Generator backbone ablation.}
    \label{tab:ablation_backbone}
    \begin{tabular}{lccc}
        \toprule
        Method & HumanEval & GSM8K & HotpotQA \\
        \midrule
        ADAS~\cite{adas} & 59.5 & 79.5 & 60.9  \\
        \method{} & \best{69.6} & \best{86.2} & \best{73.6} \\
        \midrule
        FlowMAS (VAE) & 62.6 & 81.6 & 46.1 \\
        FlowMAS (Graph Diffusion Model)
         & 55.7 & 80.8 & 50.6 \\
        FlowMAS (Transformer) & 58.8 & 82.5 & 53.2 \\
        FlowMAS (MLP) & 60.3 & 81.1 & 62.9 \\
        FlowMAS (GFlowNet, w/o CDM, w/o IOM) & 62.8 & 82.9 & 67.3 \\
        \bottomrule
    \end{tabular}
\end{table}

\paragraph{Q2: \method{}-component ablation.}
Table~\ref{tab:ablation_pid} studies the two core components of \method{}: the curiosity-driven module (CDM) and the information-guided optimization mechanism (IOM). The results show that removing either component degrades performance across all three benchmarks. Among the direct component-removal variants, removing both causes the largest drop, reducing HumanEval from 69.6 to 62.8. Replacing CDM with graph edit distance (GED)~\cite{ged} or intrinsic curiosity module (ICM)~\cite{icm} also remains worse than the full model: the ICM variant reaches 67.9 on HumanEval, only slightly above 67.3 without CDM. Replacing IOM with alternative reward designs is similarly ineffective: both Embedding Similarity and LLM-as-Judge consistently underperform the full model across all three benchmarks, with Embedding Similarity reducing HumanEval accuracy to 62.0. The topology count shows a similar trend, where the full \method{} explores more executable topologies than its component-ablated and alternative-reward variants. Overall, these results suggest that performance gains do not come from simply combining related modules; instead, the two proposed components are jointly necessary for improving both exploration diversity and workflow quality.

More ablation and analysis experiments are provided in Appendix~\ref{app:more_experiments}.

\begin{table}[!t]
    \centering
    \small
    \setlength{\tabcolsep}{5pt}
    \renewcommand{\arraystretch}{1.02}
    \caption{Q2. Method-component ablation. HE, G8K, and HQ denote accuracy on HumanEval, GSM8K, and HotpotQA, respectively. ``Avg. Topo. Count'' denotes the average number of executable topologies discovered during training across the three benchmarks.}
    \label{tab:ablation_pid}
    \resizebox{\columnwidth}{!}{%
    \begin{tabular}{lcccc}
        \toprule
        Variant & HE Acc. & G8K Acc. & HQ F1. & Avg. Topo. Count \\
        \midrule
        ADAS~\cite{adas} & 59.5 & 79.5 & 60.9 & 31 \\
        \midrule
        w/o CDM & 67.3 & 84.9 & 72.0 & 68 \\
        w/o CDM, w/ Graph Edit Distance~\cite{ged} & 66.8 & 85.2 & 72.1 & 73 \\
        w/o CDM, w/ Intrinsic Curiosity Module~\cite{icm} & 67.9 & 85.4 & 72.7 & 73 \\
        \midrule
        w/o IOM & 65.0 & 84.2 & 69.3 & 60 \\
        w/o IOM, w/ Embedding Similarity & 62.0 & 84.2 & 68.5 & 56 \\
        w/o IOM, w/ LLM-as-Judge & 65.6 & 84.9 & 69.3 & 60 \\
        w/o IOM, w/ O-Information~\cite{oinformation} & 65.0 & 84.9 & 70.6 & 66 \\
        \midrule
        w/o CDM \& IOM & 62.8 & 82.9 & 67.3 & 51 \\
        \midrule
        \method{} & \best{69.6} & \best{86.2} & \best{73.6} & 81 \\
        \bottomrule
    \end{tabular}
    }
\end{table}

\FloatBarrier

\subsection{Generalization Analysis}
\label{sec:generalization_analysis}

To further evaluate the generalization ability of \method{}, we conduct experiments on BIG-Bench Hard (BBH)~\cite{bbh_benchmark}, which covers 23 distinct task types. We sample approximately 20\% of the examples from 10 BBH task types, including Object Counting, Causal Judgement, and Formal Fallacies, to train \method{}, and evaluate on test sets totaling 5{,}000 examples from both seen and unseen task types. The unseen task types include Multistep Arithmetic, Temporal Sequences, and Logical Deduction.

As shown in Table~\ref{tab:bbh_generalization}, \method{} consistently outperforms AFlow and MaAS under both Llama-3.1-8B-Instruct and GPT-4o-mini, demonstrating its ability to generalize across diverse reasoning tasks.

\begin{table}[H]
    \centering
    \small
    \setlength{\tabcolsep}{8pt}
    \renewcommand{\arraystretch}{1.12}
    \caption{Generalization results on BIG-Bench Hard across different LLM backbones.}
    \label{tab:bbh_generalization}
    \begin{tabular}{llc}
        \toprule
        Backbone & Method & Accuracy \\
        \midrule
        \multirow{3}{*}{Llama-3.1-8B-Instruct}
        & AFlow & 43.7 \\
        & MaAS & 60.0 \\
        & \method{} & \best{69.0} \\
        \midrule
        \multirow{3}{*}{GPT-4o-mini}
        & AFlow & 75.7 \\
        & MaAS & 81.6 \\
        & \method{} & \best{85.0} \\
        \bottomrule
    \end{tabular}
\end{table}

\FloatBarrier

\subsection{Scalability with Larger Operator Spaces}
\label{sec:operator_scalability}


We also examine how \method{} scales to larger operator spaces by expanding the candidate operator set on HumanEval from 10 to 15 and 20. The added operators are deliberately chosen to be less relevant to the benchmark, including \texttt{RhetoricalCritic}, which performs rhetorical and stylistic analysis, and \texttt{CreativeStoryteller}, which generates fictional stories based on the problem statement. This setting enlarges the workflow search space and introduces distracting alternatives, testing whether \method{} can identify effective workflows from a broader, less task-focused operator pool.

As shown in Table~\ref{tab:operator_scalability}, the performance of all methods decreases as the operator space grows, while the optimization cost generally increases. Nevertheless, \method{} consistently achieves the best accuracy across different operator-space sizes and maintains a relatively low optimization cost. Even with 20 candidate operators, \method{} achieves 65.6 accuracy, compared with 58.8 for AFlow and 59.5 for MaAS.

\begin{table*}[!t]
    \centering
    \small
    \setlength{\tabcolsep}{6pt}
    \renewcommand{\arraystretch}{1.12}
    \caption{Scalability comparison on HumanEval with different numbers of candidate operators.}
    \label{tab:operator_scalability}
    \begin{tabular*}{\textwidth}{@{\extracolsep{\fill}}clcccc@{}}
        \toprule
        \# Operators & Method & Training Tokens & GPU Hours & Wall-clock Time (min) & Accuracy \\
        \midrule

        \multirow{3}{*}{10}
        & AFlow & 6{,}253{,}861 & 1.61 & 97 & 65.1 \\
        & MaAS & 223{,}042 & 0.52 & 31 & 62.7 \\
        & \method{} & 384{,}278 & 0.65 & 39 & \best{69.6} \\

        \midrule

        \multirow{3}{*}{15}
        & AFlow & 9{,}746{,}315 & 2.90 & 174 & 61.8 \\
        & MaAS & 493{,}110 & 0.78 & 47 & 61.8 \\
        & \method{} & 427{,}057 & 0.70 & 42 & \best{68.7} \\

        \midrule

        \multirow{3}{*}{20}
        & AFlow & 14{,}284{,}672 & 3.92 & 235 & 58.8 \\
        & MaAS & 627{,}634 & 0.83 & 50 & 59.5 \\
        & \method{} & 652{,}095 & 0.85 & 51 & \best{65.6} \\

        \bottomrule
    \end{tabular*}
\end{table*}
\FloatBarrier

%% file: paper_sections/related_work.tex
\section{Related Work}

Existing methods for agentic workflow topology can be roughly grouped into four categories. Static methods, such as AgentVerse~\cite{agentverse}, LLM-Debate~\cite{llm_debate}, and MetaGPT~\cite{metagpt}, rely on manually designed topologies and are therefore labor-intensive. Search-based methods treat the workflow as a design object and optimize it through repeated evaluation. Representative examples include AFlow~\cite{aflow}, which searches over topologies with tree search, ADAS~\cite{adas}, which searches in a code space, and EvoAgent~\cite{EvoAgent}, which uses population-based search. While effective, these methods usually incur high evaluation cost, and their search quality is often constrained by the workflow representation and the search strategy.

Related to search are topology pruning methods, such as DyLAN~\cite{dylan} and AgentDropout~\cite{AgentDropout}, which start from a fixed super-graph and perform run-specific activation or pruning. Although effective, their performance is bounded by the quality of the initial super-graph. Textual gradient-based methods, such as OneFlow~\cite{oneflow}, optimize workflows through text feedback from LLM-based critics or rectifiers, but such feedback is typically coarse-grained and makes structure-level credit assignment difficult.

Generation-based methods, including VAE-based G-Designer~\cite{gdesigner}, MLP-based MaAS~\cite{maas}, and diffusion-based GTD~\cite{gtd}, directly generate workflow topologies at inference time. However, these architectures are not specifically designed for discrete workflow generation with complex dependencies, and they often lack autonomous exploration. As a result, they may produce limited topology diversity and struggle to generate task-specific workflows that adapt well to different problems. In contrast, our method performs fine-grained step-level generation during training and learns a topology distribution through active exploration over the combinatorial workflow space.

Our work focuses on DAG-structured workflow topologies. Other workflow representations also exist, such as Python code in MAS-GPT~\cite{MASGPT}, domain-specific languages in AutoFlow~\cite{autoflow}, and symbolic primitive representations for executable workflow construction in AutoRAS~\cite{yue2026autoras}.

%% file: paper_sections/appendix.tex
{\centering
 \huge\bfseries Appendix\par
}

\section{FlowMAS Algorithm}

\begin{algorithm}[H]
\caption{FlowMAS}
\label{alg:masflow}
\begin{algorithmic}[1]
\State \textbf{Require:} Training set $\mathcal{D}$, operator set $\mathcal{O}$, forward-policy parameters $\theta_F$, backward-policy parameters $\theta_B$, state-flow parameters $\theta_Z$, maximum steps $T$, repetitions $N$, mini-batch $\mathcal{B}$
\State \textbf{Require:} Reward coefficients $\alpha, \beta, \lambda_{\text{cost}}$
\State Initialize graph-edit GFlowNet with forward policy $P_F(\cdot \mid s; \theta_F)$, backward policy $P_B(\cdot \mid s'; \theta_B)$, and flow normalizer parameters $\theta_Z$
\State Initialize optimizer and CDM predictor / target networks
\For{$i \gets 1$ to $N$}
    \For{each batch $\mathcal{B} \subset \mathcal{D}$ of size $|\mathcal{B}|$}
        \State Initialize an empty trajectory set $\mathcal{T} \gets \emptyset$
        \For{each problem $x \in \mathcal{B}$}
            \State Create a workflow trajectory $\tau$ with \texttt{ROOT} node
            \For{$t \gets 1$ to $T$}
                \State Encode the current state $s_t=(e_{task},e_{op},e_{topo})$
                \State Build valid graph-edit action set $\mathcal{A}(s_t)$
                \State Sample a forward action $a_t \sim P_F(\cdot \mid s_t; \theta_F)$
                \If{$\mathcal{O}_{\texttt{EarlyStop}} \in {a_t}$}
                    \State \textbf{break}
                \EndIf
                \State Execute graph edit $a_t$, transition to state $s_{t+1}$
                \State Infer the reverse edit probability $P_B(a_t^{-1} \mid s_{t+1}; \theta_B)$
                \State Compute CDM exploration reward $r_{\text{cur}}^{(t)}$
                \State $r_{\text{step}}^{(t)} \gets \alpha \cdot r_{\text{cur}}^{(t)}$
                \If{the updated workflow induces a fan-out or fan-in structure}
                    \State Compute PID-inspired synergistic information reward $r_{\text{pid}}^{(t)}$
                    \State Compute effective channel reward $r_{\text{ch}}^{(t)}$
                    \State \[
                    r_{\text{step}}^{(t)} \gets \alpha \cdot r_{\text{cur}}^{(t)} + \beta \cdot \sum_{D_{i}} (r_{\text{pid}}^{(t)} + r_{\text{ch}}^{(t)})
                    \]
                \EndIf
            \EndFor
            \State Run final verification / testing to obtain prediction $\hat{y}$ and score $s(\hat{y})$
            \State Compute terminal reward only after the full workflow topology is completed:
            \[
            R_{\text{term}} = max (0, s(\hat{y}) - \lambda_{\text{cost}} \cdot c(\hat{y})) \in [0,1]
            \]
            \State Store the trajectory $\tau$ together with  $R_{\text{term}}$ and $\{r_{\text{step}}^{(t)}\}_{t=1}^{|\tau|}$ in $\mathcal{T}$
        \EndFor
        \State Compute the GFlowNet trajectory objective $\mathcal{L}_{\text{TB}}$ over $\mathcal{T}$
        \State Update $\theta_F$, $\theta_B$, and $\theta_Z$ by backpropagating $\mathcal{L}_{\text{TB}}$
        \State Update the CDM predictor network with the visited states in $\mathcal{T}$
    \EndFor
\EndFor
\State \textbf{Return:} trained graph-edit GFlowNet $\{P_F, P_B, Z\}$
\end{algorithmic}
\end{algorithm}

\FloatBarrier


\section{Derivation of Trajectory Balance Constraint}
\label{app:TBC}
In standard GFlowNets, the flow consistency constraint is defined as: 

\begin{equation}
 F(s_{t-1})P_F(s_t \mid s_{t-1}) = F(s_t)P_B(s_{t-1} \mid s_t), 
\end{equation}

This constraint ensures that the trajectory probability from the initial state to the terminal state is consistent with the terminal reward. By introducing an intermediate reward $r(s_{t-1} \rightarrow s_{t})$ at the state transition level, we obtain a new flow consistency relation:

\begin{equation}
 F(s_{t-1})P_F(s_t \mid s_{t-1}) = F(s_t)P_B(s_{t-1} \mid s_t) + r(s_{t-1} \rightarrow s_{t}), 
 \label{equ:flow_consistency}
\end{equation}

Based on Equation~\ref{equ:flow_match_cons}, we define the forward and backward policies as:

\begin{equation}
 P_{F}(s_{t}|s_{t-1})= \frac{F(s_{t-1}\rightarrow s_{t})+r(s_{t-1}\rightarrow s_{t})}{F(s_{t-1})}, \qquad P_{B}(s_{t-1}|s_{t})=\frac{F(s_{t-1}\rightarrow s_{t})}{F(s_{t})}, 
\end{equation}

For a trajectory $\tau=\{s_{0},s_{1},...,s_{n}\}$ of length $n$, by recursively expanding Equation~\ref{equ:flow_consistency}, we have:

\begin{equation}
 \begin{aligned}
F(s_0)P_F(s_1 \mid s_0) &= F(s_1)P_B(s_0 \mid s_1) + r(s_0 \rightarrow s_1), \\
F(s_1)P_F(s_2 \mid s_1) &= F(s_2)P_B(s_1 \mid s_2) + r(s_1 \rightarrow s_2), \\
&\;\;\vdots \\
F(s_{n-1})P_F(s_n \mid s_{n-1}) &= F(s_n)P_B(s_{n-1} \mid s_n) + r(s_{n-1} \rightarrow s_n).
\end{aligned}
\end{equation}

By multiplying both sides of these equations over all transitions, we obtain:

\begin{equation}
 F(s_{0}) \cdots F(s_{n-1}) \prod_{t=0}^{n-1}P_{F}(s_{t+1}|s_{t})=F(s_{1}) \cdots F(s_{n}) \prod_{t=0}^{n-1}[P_{B}(s_{t}|s_{t+1}) + \frac{r(s_{t} \rightarrow s_{t+1})}{F(s_{t+1})}], 
\end{equation}

By rearranging the above equation, we obtain the corresponding edge-based reward-augmented formulation for trajectory balance as:

\begin{equation}
 F(s_0) \prod_{t=0}^{n-1} P_F(s_{t+1} \mid s_t) = F(s_n) \prod_{t=0}^{n-1} \left[ P_B(s_t \mid s_{t+1}) + \frac{r(s_t \rightarrow s_{t+1})}{F(s_{t+1})} \right], 
\end{equation}

where $F(s_{0})=Z_{\theta}=\sum_{x} R(x) + \sum_{s_{t-1} \rightarrow s_{t}} r(s_{t-1} \rightarrow s_{t})$, and $F(s_{n})=R(x)$. Our method relies on edge-based intermediate reward augmentation. The design details of state-based intermediate reward augmentation can be found in~\cite{gaflownets}.

\section{Detailed Calculation of the PID-inspired Reward}
\label{app:pid_reward}

This section provides the detailed computation of the PID-inspired reward introduced in Eq.~\ref{eq:pid_reward}. 
The reward is computed independently for each training instance and each local DAG primitive, without estimating mutual information from batch-level or dataset-level empirical distributions.

For a training instance, let $T$ denote the ground-truth target answer. 
Consider a local DAG primitive $D_i$ in which $K$ upstream operators $\{o_k\}_{k=1}^{K}$ produce messages $\{m_k\}_{k=1}^{K}$ that are jointly consumed by a downstream operator. 
We first encode the target answer and the incoming messages into normalized representations:
\begin{equation}
Y=
\frac{\operatorname{Emb}(T)}
     {\|\operatorname{Emb}(T)\|_2},
\qquad
X_k=
\frac{\operatorname{Emb}(m_k)}
     {\|\operatorname{Emb}(m_k)\|_2},
\end{equation}
where $\operatorname{Emb}(\cdot)$ denotes the embedding encoder.

We measure the target alignment of each incoming message as
\begin{equation}
I(X_k;Y)
=
\max{(0,\operatorname{sim}(X_k,Y))},
\end{equation}
where $\operatorname{sim}(\cdot,\cdot)$ denotes cosine similarity.

To characterize the target alignment of the aggregated incoming messages, we compute
\begin{equation}
I(D_i;Y)
=
\max{(0,
\operatorname{sim}
\left(
\frac{\sum_{k=1}^{K}X_k}
     {\left\|\sum_{k=1}^{K}X_k\right\|_2},
Y
\right)
)}.
\end{equation}

Following the redundancy--uniqueness--synergy intuition of Partial Information Decomposition (PID), we approximate the target-aligned information shared across all incoming messages as
\begin{equation}
Red(D_i;Y)
=
\min_{k} I(X_k;Y).
\end{equation}
The contribution unique to the $k$-th message is then defined as
\begin{equation}
UI_k(D_i;Y)
=
I(X_k;Y)-Red(D_i;Y).
\end{equation}

Finally, the PID-inspired synergistic information reward is computed as
\begin{equation}
r_{pid}(D_i)
=
\max{(0,
I(D_i;Y)
-
\sum_{k=1}^{K} UI_k(D_i;Y)
-
Red(D_i;Y)
 )} \in {[0, 1]}.
\end{equation}

Notably, the target representation $Y$ and the resulting reward are used only during training; neither the ground-truth answer nor the IOM reward is required during inference.

\section{Derivation of Spectral Entropy}
\label{app:SE}

For a local DAG primitive $D_i$, the outputs of its $K$ upstream operators are denoted as
$\{Z_{1},Z_{2},\ldots,Z_{K}\}$. 
For each output, the normalized embedding is computed as

\begin{equation}
\tilde{z}_{j}
:=
\frac{\operatorname{Emb}(Z_{j})}
     {\|\operatorname{Emb}(Z_{j})\|_{2}}
\in \mathbb{R}^{d},
\end{equation}

where $\operatorname{Emb}(\cdot)$ denotes the embedding model. 
Further, the cosine-similarity Gram matrix $\mathcal{M}$ is computed as

\begin{equation}
\mathcal{M}
\in
\mathbb{R}^{K \times K},
\qquad
\mathcal{M}_{pq}
=
\tilde{z}_{p}^{\top}\tilde{z}_{q}.
\end{equation}

Applying trace normalization to the Gram matrix $\mathcal{M}$ yields

\begin{equation}
\rho
=
\frac{\mathcal{M}}
     {\operatorname{Tr}(\mathcal{M})}.
\end{equation}

Let $\{\lambda_{1},\lambda_{2},\ldots,\lambda_{K}\}$ denote the eigenvalues of $\rho$. 
Based on the eigenvalue distribution, the spectral entropy $H(\rho)$ is defined as

\begin{equation}
H(\rho)
=
-\sum_{j=1}^{K}
\lambda_{j}\log_{2}\lambda_{j}.
\end{equation}

Accordingly, the effective channel reward is computed as

\begin{equation}
r_{ch}(D_i)
=
\frac{2^{H(\rho)}-1}{K-1}.
\end{equation}

\section{Case Study}
\label{sec:case_study}

\begin{figure}[H]
    \centering
    \includegraphics[width=\textwidth]{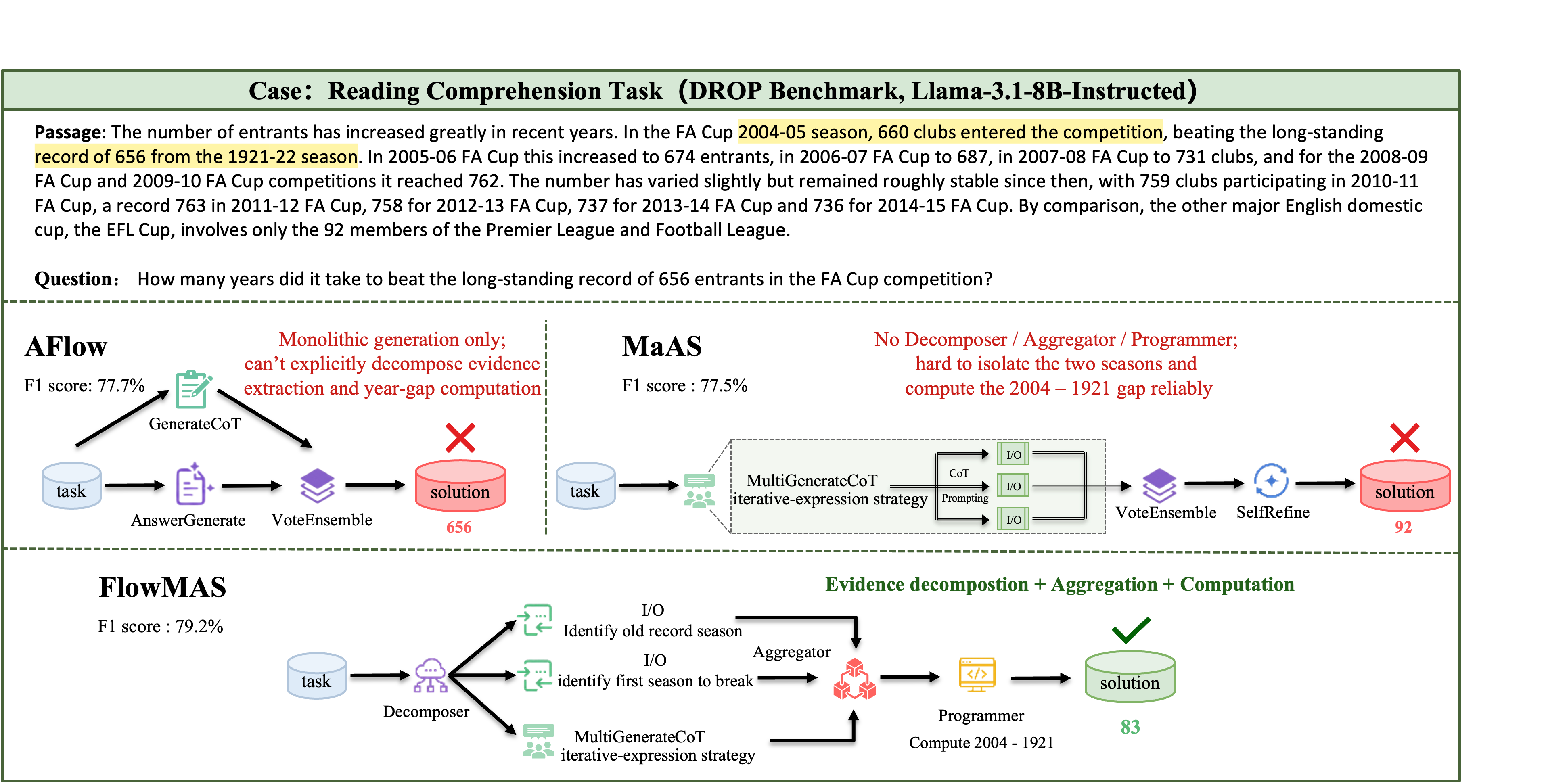}
    \caption{Case study of the workflow topologies designed by AFlow, MaAS, and \method{} on the DROP benchmark.}
    \label{fig:case_study}
\end{figure}

Figure~\ref{fig:case_study} visualizes the different workflow topologies designed by AFlow, MaAS, and \method{} on the DROP benchmark. Compared with AFlow and MaAS, \method{} produces a more task-adaptive workflow structure that better separates evidence extraction, reasoning, and answer aggregation.

\section{Baseline Methods and FlowMAS Variants}
\label{app:baseline_methods_detail}

We consider three categories of methods for comparison: (1) single-agent test-time scaling methods: Vanilla LLM (IO), CoT, CoT-SC, ReAct, and Reflexion; (2) hand-crafted multi-agent systems: LLM-Majority, LLM-Debate, LLM-Blender, DyLAN, and AgentVerse; (3) automated multi-agent systems: ADAS, AgentNet, SELFORG, AFlow, and MaAS.

\begin{enumerate}
\item \textbf{Vanilla LLM (IO)}: a direct input-output baseline that queries the LLM once and uses the returned answer without explicit deliberation, search, or multi-agent coordination.
\item \textbf{CoT~\cite{cot}}: a prompt-based method that guides the LLM to generate intermediate reasoning steps before producing the final answer.
\item \textbf{CoT-SC~\cite{sc}}: a test-time scaling method that samples multiple chain-of-thought solutions and selects the final answer by majority voting across sampled outputs.
\item \textbf{ReAct~\cite{react}}: a prompting framework that interleaves reasoning and action generation, allowing the model to iteratively reason over observations and external feedback\footnotemark[1].
\item \textbf{Reflexion~\cite{reflexion}}: a prompt-based method that converts verbalized environmental feedback into self-reflection to improve subsequent attempts\footnotemark[2].
\item \textbf{LLM-Majority~\cite{llm_majority}}: a simple multi-agent baseline that queries multiple LLM instances independently and uses majority voting to obtain the final prediction.
\item \textbf{LLM-Debate~\cite{llm_debate}}: a multi-agent deliberation method in which several LLM agents iteratively debate and refine their positions before reaching a final answer\footnotemark[3].
\item \textbf{LLM-Blender~\cite{llm_blender}}: an ensemble framework that combines and reranks candidate outputs from multiple LLMs or decoding paths to produce a stronger final response\footnotemark[4].
\item \textbf{DyLAN~\cite{dylan}}: a dynamic multi-agent system that adaptively invokes different agents and interaction patterns according to the evolving problem-solving state\footnotemark[5].
\item \textbf{AgentVerse~\cite{agentverse}}: a configurable multi-agent platform that organizes role-based agents into predefined collaboration workflows for task solving\footnotemark[6].
\item \textbf{ADAS~\cite{adas}}: an automated design framework that utilizes a search algorithm and a meta-agent to iteratively invent, program, and optimize novel agentic building blocks and workflows directly in code\footnotemark[7].
\item \textbf{AgentNet~\cite{agentnet}}: a decentralized, RAG-based framework that enables LLM-based agents to specialize, evolve, and collaborate autonomously in a dynamically structured Directed Acyclic Graph\footnotemark[8]
\item \textbf{SELFORG~\cite{selforg}}: a decentralized multi-agent framework that dynamically constructs an adaptive communication graph on-the-fly by evaluating peer responses using Shapley-based contribution scores, routing information from high-contributing agents to others without relying on external judges or fixed topologies\footnotemark[9].
\item \textbf{AFlow~\cite{aflow}}: an automated framework that leverages Monte Carlo Tree Search and LLMs as optimizers to systematically explore, iteratively refine, and discover optimal code-represented agentic workflows using execution feedback\footnotemark[10].
\item \textbf{MaAS~\cite{maas}}: an automated framework that optimizes a probabilistic agentic supernet to dynamically sample query-dependent multi-agent architectures, delivering tailored solutions and efficient resource allocation\footnotemark[11].
\end{enumerate}

\footnotetext[1]{ReAct: \url{https://github.com/ysymyth/ReAct}}
\footnotetext[2]{Reflexion: \url{https://github.com/noahshinn/reflexion}}
\footnotetext[3]{LLM-Debate: \url{https://github.com/ucl-dark/llm_debate}}
\footnotetext[4]{LLM-Blender: \url{https://github.com/yuchenlin/LLM-Blender}}
\footnotetext[5]{DyLAN: \url{https://github.com/SALT-NLP/DyLAN}}
\footnotetext[6]{AgentVerse: \url{https://github.com/openbmb/agentverse}}
\footnotetext[7]{ADAS: \url{https://github.com/ShengranHu/ADAS}}
\footnotetext[8]{AgentNet: \url{https://github.com/zoe-yyx/AgentNet}}
\footnotetext[9]{SELFORG: \url{https://github.com/tnurbek/selforg}}
\footnotetext[10]{AFlow: \url{https://github.com/FoundationAgents/AFlow}}
\footnotetext[11]{MaAS: \url{https://github.com/bingreeky/MaAS}}

In this work, we further consider several \method{} variants to evaluate the contribution of different design choices. In particular, we include these variants to compare different exploration-encouraging strategies~\cite{rnd, icm, ged} and information decomposition mechanisms~\cite{williams2010nonnegative, oinformation}. These variants are listed as follows.

\begin{enumerate}
\item \textbf{FlowMAS w/o IOM}: a variant of FlowMAS, where the information-guided optimization mechanism (IOM) is removed.
\item \textbf{FlowMAS w/o CDM}: a variant of FlowMAS, where the curiosity-driven module (CDM) is removed.
\item \textbf{FlowMAS w/o IOM, w/o CDM}: a variant of FlowMAS, where the information-guided optimization mechanism (IOM) and the curiosity-driven module (CDM) are removed.
\item \textbf{FlowMAS w/o IOM, w/ O-Information~\cite{oinformation}}: a variant of FlowMAS, where the information-guided optimization mechanism (IOM) is removed and replaced with an O-Information-based reward.
\item \textbf{FlowMAS w/o CDM, w/ Graph Edit Distance~\cite{ged}}: a variant of FlowMAS, where the curiosity-driven module (CDM) is removed and replaced with a Graph Edit Distance-based diversity reward.
\item \textbf{FlowMAS w/o CDM, w/ Intrinsic Curiosity Module~\cite{icm}}: a variant of FlowMAS, where the curiosity-driven module (CDM) is removed and replaced with an Intrinsic Curiosity Module-based exploration reward.
\item \textbf{FlowMAS w/o IOM, w/ Embedding Similarity}: a variant of FlowMAS, where the information-guided optimization mechanism (IOM) is removed and replaced with a direct embedding similarity-based reward. Specifically, given the normalized representations $\{X_k\}_{k=1}^{K}$ of the incoming messages and the target representation $Y$, the reward is defined as
\begin{equation}
r_{\text{emb-sim}}
=
\frac{
1+
\operatorname{sim}
\left(
\frac{\sum_{k=1}^{K}X_k}
     {\left\|\sum_{k=1}^{K}X_k\right\|_2},
Y
\right)
}{2}
\in [0,1],
\end{equation}
where $\operatorname{sim}(\cdot,\cdot)$ denotes cosine similarity. This reward directly measures the similarity between the aggregated incoming messages and the target representation.

\item \textbf{FlowMAS w/o IOM, w/ LLM-as-Judge}: a variant of FlowMAS, where the information-guided optimization mechanism (IOM) is removed and replaced with an LLM-as-Judge reward. Specifically, an LLM judge is prompted to estimate the information reward captured by IOM for each intermediate collaboration pattern.

\end{enumerate}

In addition, we include the designations used for various ablation experiments.

\begin{enumerate}
\item \textbf{Graph Diffusion Model~\cite{DDPM, xu2024discrete} (ref. GTD~\cite{gtd})}: following GTD, we use a Graph Diffusion Model as the basic network backbone for multi-agent system generation, with other method-specific optimizations removed\footnotemark[12]. 
\item \textbf{VAE~\cite{VAE} (ref. G-Designer~\cite{gdesigner})}: following G-Designer, we use a VAE as the basic network backbone for multi-agent system generation, with other method-specific optimizations removed\footnotemark[13]. 
\item \textbf{MLP~\cite{mlp} (ref. MaAS~\cite{maas})}: following MaAS, we use a multi-layer MLP as the basic network backbone for multi-agent system generation, with other method-specific optimizations removed\footnotemark[11]. 
\item \textbf{Transformer~\cite{transformer} (ref. GTD~\cite{gtd})}: we use a Transformer as the basic network backbone for multi-agent system generation. The supervision data are constructed following GTD\footnotemark[12].
\end{enumerate}

\footnotetext[12]{GTD: \url{https://github.com/ericjiang18/diffusion_agent}} 
\footnotetext[13]{G-Designer: \url{https://github.com/yanweiyue/GDesigner}}

\section{Benchmark Detail and More Experimental}
\label{app:more_experiments}

\subsection{Benchmark Statistics}
\label{app:benchmark_statistics}

Following established data partition protocols used in prior workflow-automation studies, we split each benchmark into a training/validation portion and a test portion with a 1:4 ratio. For MATH, we select 605 difficulty-level-5 problems from four representative categories: combinatorics and probability, number theory, pre-algebra, and pre-calculus. For HotpotQA and DROP, we randomly sample 1{,}000 examples from each benchmark following ADAS~\cite{adas}. The resulting dataset statistics are summarized in Table~\ref{tab:benchmark_statistics}.

\begin{table}[!t]
    \centering
    \small
    \setlength{\tabcolsep}{5pt}
    \renewcommand{\arraystretch}{1.12}
    \caption{Dataset statistics used in our experiments.}
    \label{tab:benchmark_statistics}
    \begin{tabular*}{\linewidth}{@{\extracolsep{\fill}}llccc@{}}
        \toprule
        Domain & Dataset & \#Train/Val & \#Test & Metric \\
        \midrule
        \multirow{2}{*}{\shortstack[l]{Code\\Generation}} & HumanEval & 33 & 131 & pass@1 \\
         & MBPP & 86 & 341 & pass@1 \\
        \midrule
        \multirow{2}{*}{\shortstack[l]{Math\\Reasoning}} & GSM8K & 264 & 1{,}055 & Solve rate \\
         & MATH & 119 & 486 & Solve rate \\
        \midrule
        \multirow{2}{*}{\shortstack[l]{Reading\\Comprehension}} & HotpotQA & 200 & 800 & F1 \\
         & DROP & 200 & 800 & F1 \\
        \bottomrule
    \end{tabular*}
\end{table}

\subsection{Supplementary Experiments}
\label{app:supplementary_experiments}

\subsubsection{Additional Results on More LLM Backbones}
\label{app:additional_backbone_results}

Detailed results of all methods on Gemma-4-26B-A4B-it and GPT-4o-mini are reported in Table~\ref{tab:appendix_gemma26b} and Table~\ref{tab:appendix_gpt4omini}, respectively.

\begin{table*}[!t]
    \centering
    \renewcommand{\arraystretch}{1.12}
    \caption{Detailed results on GPT-4o-mini. The best result in each column is boldfaced, and the second-best is underlined. Each experiment is run five times to record the mean, standard deviation.}
    \label{tab:appendix_gpt4omini}
    \resizebox{\textwidth}{!}{%
    \begin{tabular}{l|ccccccc}
        \toprule
        \rowcolor{black!4}
        Method & HumanEval & GSM8K & MATH & MBPP & HotpotQA & DROP & Avg. \\
        \midrule
        \rowcolor{black!4}
        IO & 86.3\textcolor{SlateBlue}{\scriptsize$\pm 0.6$} & 86.3\textcolor{SlateBlue}{\scriptsize$\pm 0.5$} & 45.0\textcolor{SlateBlue}{\scriptsize$\pm 0.7$} & 71.2\textcolor{SlateBlue}{\scriptsize$\pm 0.4$} & 66.7\textcolor{SlateBlue}{\scriptsize$\pm 0.6$} & 73.2\textcolor{SlateBlue}{\scriptsize$\pm 0.9$} & 71.45 \\
        CoT~\cite{cot} & 86.6\textcolor{SlateBlue}{\scriptsize$\pm 0.7$} & 86.4\textcolor{SlateBlue}{\scriptsize$\pm 0.5$} & 45.8\textcolor{SlateBlue}{\scriptsize$\pm 0.6$} & 72.6\textcolor{SlateBlue}{\scriptsize$\pm 0.6$} & 66.4\textcolor{SlateBlue}{\scriptsize$\pm 0.9$} & 77.0\textcolor{SlateBlue}{\scriptsize$\pm 0.8$} & 72.47 \\
        \rowcolor{black!4}
        CoT-SC~\cite{sc} & 87.6\textcolor{SlateBlue}{\scriptsize$\pm 0.0$} & 86.5\textcolor{SlateBlue}{\scriptsize$\pm 0.3$} & 46.5\textcolor{SlateBlue}{\scriptsize$\pm 0.0$} & 74.2\textcolor{SlateBlue}{\scriptsize$\pm 0.5$} & 67.4\textcolor{SlateBlue}{\scriptsize$\pm 0.6$} & 77.5\textcolor{SlateBlue}{\scriptsize$\pm 0.7$} & 73.28 \\
        ReAct~\cite{react} & 87.0\textcolor{SlateBlue}{\scriptsize$\pm 0.8$} & 86.5\textcolor{SlateBlue}{\scriptsize$\pm 0.7$} & 43.3\textcolor{SlateBlue}{\scriptsize$\pm 0.9$} & 77.7\textcolor{SlateBlue}{\scriptsize$\pm 0.6$} & 71.0\textcolor{SlateBlue}{\scriptsize$\pm 0.8$} & 73.8\textcolor{SlateBlue}{\scriptsize$\pm 0.9$} & 73.22 \\
        \rowcolor{black!4}
        Reflexion~\cite{reflexion} & 88.7\textcolor{SlateBlue}{\scriptsize$\pm 0.7$} & 87.5\textcolor{SlateBlue}{\scriptsize$\pm 0.4$} & 45.9\textcolor{SlateBlue}{\scriptsize$\pm 0.9$} & 76.8\textcolor{SlateBlue}{\scriptsize$\pm 0.6$} & 66.2\textcolor{SlateBlue}{\scriptsize$\pm 0.7$} & 77.1\textcolor{SlateBlue}{\scriptsize$\pm 0.8$} & 73.70 \\
        \groupsep
        LLM-Majority~\cite{llm_majority} & 87.9\textcolor{SlateBlue}{\scriptsize$\pm 0.3$} & 86.5\textcolor{SlateBlue}{\scriptsize$\pm 0.6$} & 46.8\textcolor{SlateBlue}{\scriptsize$\pm 0.0$} & 74.4\textcolor{SlateBlue}{\scriptsize$\pm 0.0$} & 68.3\textcolor{SlateBlue}{\scriptsize$\pm 0.6$} & 74.6\textcolor{SlateBlue}{\scriptsize$\pm 0.5$} & 73.08 \\
        \rowcolor{black!4}
        LLM-Debate~\cite{llm_debate} & 87.9\textcolor{SlateBlue}{\scriptsize$\pm 0.6$} & 88.2\textcolor{SlateBlue}{\scriptsize$\pm 0.9$} & 47.8\textcolor{SlateBlue}{\scriptsize$\pm 0.8$} & 72.1\textcolor{SlateBlue}{\scriptsize$\pm 0.5$} & 65.5\textcolor{SlateBlue}{\scriptsize$\pm 0.9$} & 71.9\textcolor{SlateBlue}{\scriptsize$\pm 0.6$} & 72.23 \\
        LLM-Blender~\cite{llm_blender} & 87.9\textcolor{SlateBlue}{\scriptsize$\pm 0.8$} & 87.5\textcolor{SlateBlue}{\scriptsize$\pm 0.9$} & 46.2\textcolor{SlateBlue}{\scriptsize$\pm 0.8$} & 75.6\textcolor{SlateBlue}{\scriptsize$\pm 0.6$} & 68.1\textcolor{SlateBlue}{\scriptsize$\pm 0.9$} & 73.9\textcolor{SlateBlue}{\scriptsize$\pm 0.7$} & 73.20 \\
        \rowcolor{black!4}
        DyLAN~\cite{dylan} & 89.6\textcolor{SlateBlue}{\scriptsize$\pm 0.7$} & 88.9\textcolor{SlateBlue}{\scriptsize$\pm 0.9$} & 47.1\textcolor{SlateBlue}{\scriptsize$\pm 1.2$} & 78.1\textcolor{SlateBlue}{\scriptsize$\pm 0.7$} & 70.4\textcolor{SlateBlue}{\scriptsize$\pm 0.9$} & 76.2\textcolor{SlateBlue}{\scriptsize$\pm 0.8$} & 75.05 \\
        AgentVerse~\cite{agentverse} & 88.3\textcolor{SlateBlue}{\scriptsize$\pm 0.9$} & 89.4\textcolor{SlateBlue}{\scriptsize$\pm 1.2$} & 46.3\textcolor{SlateBlue}{\scriptsize$\pm 0.9$} & 78.0\textcolor{SlateBlue}{\scriptsize$\pm 0.7$} & 68.2\textcolor{SlateBlue}{\scriptsize$\pm 0.6$} & 75.4\textcolor{SlateBlue}{\scriptsize$\pm 0.8$} & 74.27 \\
        \groupsep
        \rowcolor{black!4}
        ADAS~\cite{adas} & 83.4\textcolor{SlateBlue}{\scriptsize$\pm 1.2$} & 84.9\textcolor{SlateBlue}{\scriptsize$\pm 0.9$} & 42.7\textcolor{SlateBlue}{\scriptsize$\pm 1.2$} & 67.3\textcolor{SlateBlue}{\scriptsize$\pm 0.8$} & 63.9\textcolor{SlateBlue}{\scriptsize$\pm 0.9$} & 75.5\textcolor{SlateBlue}{\scriptsize$\pm 1.2$} & 69.62 \\
        AgentNet~\cite{agentnet} & 86.8\textcolor{SlateBlue}{\scriptsize$\pm 0.9$} & 91.8\textcolor{SlateBlue}{\scriptsize$\pm 0.0$} & 47.5\textcolor{SlateBlue}{\scriptsize$\pm 0.1$} & 71.1\textcolor{SlateBlue}{\scriptsize$\pm 0.2$} & 64.5\textcolor{SlateBlue}{\scriptsize$\pm 0.2$} & 80.8\textcolor{SlateBlue}{\scriptsize$\pm 0.0$} & 73.74 \\
        \rowcolor{black!4}
        SELFORG~\cite{selforg} & 86.0\textcolor{SlateBlue}{\scriptsize$\pm 2.0$} & \second{92.7\textcolor{SlateBlue}{\scriptsize$\pm 0.1$}} & 50.7\textcolor{SlateBlue}{\scriptsize$\pm 0.7$} & 71.7\textcolor{SlateBlue}{\scriptsize$\pm 0.3$} & 68.9\textcolor{SlateBlue}{\scriptsize$\pm 0.0$} & \second{84.3\textcolor{SlateBlue}{\scriptsize$\pm 0.1$}} & 75.71 \\
        AFlow~\cite{aflow} & 89.6\textcolor{SlateBlue}{\scriptsize$\pm 0.7$} & 90.5\textcolor{SlateBlue}{\scriptsize$\pm 0.7$} & 50.5\textcolor{SlateBlue}{\scriptsize$\pm 0.9$} & 81.0\textcolor{SlateBlue}{\scriptsize$\pm 0.6$} & 72.6\textcolor{SlateBlue}{\scriptsize$\pm 0.9$} & 79.7\textcolor{SlateBlue}{\scriptsize$\pm 0.5$} & 77.32 \\
        \rowcolor{black!4}
        MaAS~\cite{maas} & \second{92.2\textcolor{SlateBlue}{\scriptsize$\pm 0.5$}} & 91.7\textcolor{SlateBlue}{\scriptsize$\pm 0.6$} & \second{51.1\textcolor{SlateBlue}{\scriptsize$\pm 0.8$}} & \second{81.4\textcolor{SlateBlue}{\scriptsize$\pm 0.5$}} & \second{77.3\textcolor{SlateBlue}{\scriptsize$\pm 0.7$}} & 83.1\textcolor{SlateBlue}{\scriptsize$\pm 0.6$} & \second{79.47} \\
        \groupsep
        \rowcolor{methodhighlight}
        \method{} & \best{94.4\textcolor{SlateBlue}{\scriptsize$\pm 0.8$}} & \best{93.1\textcolor{SlateBlue}{\scriptsize$\pm 0.4$}} & \best{52.6\textcolor{SlateBlue}{\scriptsize$\pm 0.7$}} & \best{83.4\textcolor{SlateBlue}{\scriptsize$\pm 0.5$}} & \best{80.0\textcolor{SlateBlue}{\scriptsize$\pm 0.4$}} & \best{84.9\textcolor{SlateBlue}{\scriptsize$\pm 0.2$}} & \best{81.40} \\
        \bottomrule
    \end{tabular}%
    }
\end{table*}

\begin{table*}[!t]
    \centering
    \renewcommand{\arraystretch}{1.12}
    \caption{Detailed results on Gemma-4-26B-A4B-it. The best result in each column is boldfaced, and the second-best is underlined. Each experiment is run five times to record the mean, standard deviation.}
    \label{tab:appendix_gemma26b}
    \resizebox{\textwidth}{!}{%
    \begin{tabular}{l|ccccccc}
        \toprule
        \rowcolor{black!4}
        Method & HumanEval & GSM8K & MATH & MBPP & HotpotQA & DROP & Avg. \\
        \midrule
        \rowcolor{black!4}
        IO & 90.8\textcolor{SlateBlue}{\scriptsize$\pm 0.8$} & 90.5\textcolor{SlateBlue}{\scriptsize$\pm 0.7$} & 61.3\textcolor{SlateBlue}{\scriptsize$\pm 0.9$} & 73.3\textcolor{SlateBlue}{\scriptsize$\pm 0.6$} & 74.9\textcolor{SlateBlue}{\scriptsize$\pm 0.8$} & 83.9\textcolor{SlateBlue}{\scriptsize$\pm 1.0$} & 79.12 \\
        CoT~\cite{cot} & 91.6\textcolor{SlateBlue}{\scriptsize$\pm 0.9$} & 91.6\textcolor{SlateBlue}{\scriptsize$\pm 0.6$} & 61.7\textcolor{SlateBlue}{\scriptsize$\pm 1.0$} & 76.7\textcolor{SlateBlue}{\scriptsize$\pm 0.7$} & 79.3\textcolor{SlateBlue}{\scriptsize$\pm 1.1$} & 87.4\textcolor{SlateBlue}{\scriptsize$\pm 0.9$} & 81.38 \\
        \rowcolor{black!4}
        CoT-SC~\cite{sc} & 87.6\textcolor{SlateBlue}{\scriptsize$\pm 0.0$} & 91.2\textcolor{SlateBlue}{\scriptsize$\pm 0.5$} & 61.7\textcolor{SlateBlue}{\scriptsize$\pm 0.3$} & 78.0\textcolor{SlateBlue}{\scriptsize$\pm 0.4$} & 78.3\textcolor{SlateBlue}{\scriptsize$\pm 0.6$} & 87.8\textcolor{SlateBlue}{\scriptsize$\pm 0.7$} & 80.77 \\
        ReAct~\cite{react} & 90.0\textcolor{SlateBlue}{\scriptsize$\pm 1.0$} & 90.3\textcolor{SlateBlue}{\scriptsize$\pm 0.9$} & 60.0\textcolor{SlateBlue}{\scriptsize$\pm 1.1$} & 77.4\textcolor{SlateBlue}{\scriptsize$\pm 0.8$} & 78.6\textcolor{SlateBlue}{\scriptsize$\pm 1.0$} & 88.2\textcolor{SlateBlue}{\scriptsize$\pm 0.9$} & 80.75 \\
        \rowcolor{black!4}
        Reflexion~\cite{reflexion} & 91.8\textcolor{SlateBlue}{\scriptsize$\pm 0.9$} & 92.1\textcolor{SlateBlue}{\scriptsize$\pm 0.6$} & 62.2\textcolor{SlateBlue}{\scriptsize$\pm 1.2$} & 76.2\textcolor{SlateBlue}{\scriptsize$\pm 0.7$} & 73.5\textcolor{SlateBlue}{\scriptsize$\pm 0.8$} & 85.1\textcolor{SlateBlue}{\scriptsize$\pm 1.0$} & 80.15 \\
        \groupsep
        LLM-Majority~\cite{llm_majority} & 91.7\textcolor{SlateBlue}{\scriptsize$\pm 0.8$} & 91.2\textcolor{SlateBlue}{\scriptsize$\pm 0.4$} & 61.3\textcolor{SlateBlue}{\scriptsize$\pm 0.4$} & 78.5\textcolor{SlateBlue}{\scriptsize$\pm 0.0$} & 79.3\textcolor{SlateBlue}{\scriptsize$\pm 0.7$} & 88.2\textcolor{SlateBlue}{\scriptsize$\pm 0.6$} & 81.70 \\
        \rowcolor{black!4}
        LLM-Debate~\cite{llm_debate} & 92.3\textcolor{SlateBlue}{\scriptsize$\pm 0.8$} & 89.3\textcolor{SlateBlue}{\scriptsize$\pm 0.9$} & 56.9\textcolor{SlateBlue}{\scriptsize$\pm 1.0$} & 75.3\textcolor{SlateBlue}{\scriptsize$\pm 0.6$} & 76.6\textcolor{SlateBlue}{\scriptsize$\pm 0.9$} & 84.3\textcolor{SlateBlue}{\scriptsize$\pm 0.8$} & 79.12 \\
        LLM-Blender~\cite{llm_blender} & 89.9\textcolor{SlateBlue}{\scriptsize$\pm 0.9$} & 90.1\textcolor{SlateBlue}{\scriptsize$\pm 1.1$} & 61.4\textcolor{SlateBlue}{\scriptsize$\pm 1.0$} & 77.1\textcolor{SlateBlue}{\scriptsize$\pm 0.7$} & 77.7\textcolor{SlateBlue}{\scriptsize$\pm 1.0$} & 87.6\textcolor{SlateBlue}{\scriptsize$\pm 0.8$} & 80.63 \\
        \rowcolor{black!4}
        DyLAN~\cite{dylan} & 92.3\textcolor{SlateBlue}{\scriptsize$\pm 0.8$} & 91.8\textcolor{SlateBlue}{\scriptsize$\pm 1.0$} & 62.7\textcolor{SlateBlue}{\scriptsize$\pm 1.1$} & 79.5\textcolor{SlateBlue}{\scriptsize$\pm 0.8$} & 70.3\textcolor{SlateBlue}{\scriptsize$\pm 1.0$} & 88.2\textcolor{SlateBlue}{\scriptsize$\pm 0.9$} & 80.80 \\
        AgentVerse~\cite{agentverse} & 89.8\textcolor{SlateBlue}{\scriptsize$\pm 1.0$} & 92.6\textcolor{SlateBlue}{\scriptsize$\pm 1.2$} & 62.0\textcolor{SlateBlue}{\scriptsize$\pm 1.1$} & 79.2\textcolor{SlateBlue}{\scriptsize$\pm 0.9$} & 78.8\textcolor{SlateBlue}{\scriptsize$\pm 0.8$} & 86.6\textcolor{SlateBlue}{\scriptsize$\pm 0.9$} & 81.50 \\
        \groupsep
        \rowcolor{black!4}
        ADAS~\cite{adas} & 88.5\textcolor{SlateBlue}{\scriptsize$\pm 1.2$} & 90.0\textcolor{SlateBlue}{\scriptsize$\pm 1.0$} & 59.2\textcolor{SlateBlue}{\scriptsize$\pm 1.2$} & 71.2\textcolor{SlateBlue}{\scriptsize$\pm 0.8$} & 76.7\textcolor{SlateBlue}{\scriptsize$\pm 1.1$} & 86.5\textcolor{SlateBlue}{\scriptsize$\pm 1.2$} & 78.68 \\
        AgentNet~\cite{agentnet} & 90.9\textcolor{SlateBlue}{\scriptsize$\pm 1.2$} & 91.9\textcolor{SlateBlue}{\scriptsize$\pm 1.0$} & 62.7\textcolor{SlateBlue}{\scriptsize$\pm 0.3$} & 81.6\textcolor{SlateBlue}{\scriptsize$\pm 0.3$} & 79.5\textcolor{SlateBlue}{\scriptsize$\pm 0.4$} & 88.8\textcolor{SlateBlue}{\scriptsize$\pm 1.0$} & 82.57 \\
        \rowcolor{black!4}
        SELFORG~\cite{selforg} & 91.3\textcolor{SlateBlue}{\scriptsize$\pm 0.3$} & 92.1\textcolor{SlateBlue}{\scriptsize$\pm 0.3$} & 63.1\textcolor{SlateBlue}{\scriptsize$\pm 0.4$} & 77.8\textcolor{SlateBlue}{\scriptsize$\pm 0.4$} & 79.5\textcolor{SlateBlue}{\scriptsize$\pm 0.5$} & 90.4\textcolor{SlateBlue}{\scriptsize$\pm 0.5$} & 82.37 \\
        AFlow~\cite{aflow} & \second{92.6\textcolor{SlateBlue}{\scriptsize$\pm 0.8$}} & 93.4\textcolor{SlateBlue}{\scriptsize$\pm 0.9$} & 65.0\textcolor{SlateBlue}{\scriptsize$\pm 1.0$} & \second{83.5\textcolor{SlateBlue}{\scriptsize$\pm 0.7$}} & 82.7\textcolor{SlateBlue}{\scriptsize$\pm 1.0$} & 89.9\textcolor{SlateBlue}{\scriptsize$\pm 1.1$} & 84.52 \\
        \rowcolor{black!4}
        MaAS~\cite{maas} & 92.4\textcolor{SlateBlue}{\scriptsize$\pm 0.6$} & \second{95.1\textcolor{SlateBlue}{\scriptsize$\pm 0.7$}} & \second{65.1\textcolor{SlateBlue}{\scriptsize$\pm 1.0$}} & 83.1\textcolor{SlateBlue}{\scriptsize$\pm 0.6$} & \second{83.8\textcolor{SlateBlue}{\scriptsize$\pm 0.8$}} & \second{90.7\textcolor{SlateBlue}{\scriptsize$\pm 0.7$}} & \second{85.03} \\
        \groupsep
        \rowcolor{methodhighlight}
        \method{} & \best{94.8\textcolor{SlateBlue}{\scriptsize$\pm 0.8$}} & \best{95.6\textcolor{SlateBlue}{\scriptsize$\pm 0.8$}} & \best{66.7\textcolor{SlateBlue}{\scriptsize$\pm 0.9$}} & \best{87.1\textcolor{SlateBlue}{\scriptsize$\pm 0.6$}} & \best{85.3\textcolor{SlateBlue}{\scriptsize$\pm 0.8$}} & \best{92.0\textcolor{SlateBlue}{\scriptsize$\pm 0.7$}} & \best{86.92} \\
        \bottomrule
    \end{tabular}%
    }
\end{table*}

%



\subsubsection{Hyperparameter Sensitivity}
\label{app:hyperparameter_sensitivity}

We further analyze the sensitivity of the reward coefficients $\alpha$ and $\beta$ in Eq.~\ref{equ:reward}, which control the contributions of CDM and IOM, respectively. The experiments are conducted on HumanEval with Llama-3.1-8B-Instruct. When varying $\alpha$, we fix $\beta=0.8$; when varying $\beta$, we fix $\alpha=0.2$.

As shown in Table~\ref{tab:hyperparameter_sensitivity}, \method{} remains relatively stable across the evaluated hyperparameter ranges. We therefore use $\alpha=0.2$ and $\beta=0.8$ as the default configuration in our experiments.

\begin{table}[!t]
    \centering
    \small
    \setlength{\tabcolsep}{7pt}
    \renewcommand{\arraystretch}{1.12}
    \caption{Hyperparameter sensitivity analysis on HumanEval with Llama-3.1-8B-Instruct.}
    \label{tab:hyperparameter_sensitivity}
    \begin{tabular}{lccccc}
        \toprule
        $\alpha$ & 0.1 & 0.2 & 0.3 & 0.4 & 0.5 \\
        \midrule
        Accuracy & 67.9 & \best{69.6} & \best{69.6} & 68.7 & 67.9 \\
        \midrule
        $\beta$ & 0.6 & 0.7 & 0.8 & 0.9 & 1.0 \\
        \midrule
        Accuracy & 66.4 & 68.7 & \best{69.6} & 67.9 & 67.9 \\
        \bottomrule
    \end{tabular}
\end{table}

\subsubsection{Training and Inference Wall-clock Time}
\label{app:wallclock_time}

In addition to token consumption, we report the end-to-end wall-clock time of representative automated workflow optimization methods on HumanEval. All experiments are measured on a single NVIDIA GeForce RTX 5090 with Llama-3.1-8B-Instruct. The reported time includes the complete optimization or inference procedure of each method.

As shown in Table~\ref{tab:wallclock_time}, \method{} requires substantially less training time than AFlow and ADAS while achieving the shortest inference time among the compared methods. Although MaAS requires slightly less training time, \method{} provides a better trade-off between optimization cost and task performance.

\begin{table}[!t]
    \centering
    \small
    \setlength{\tabcolsep}{8pt}
    \renewcommand{\arraystretch}{1.12}
    \caption{End-to-end training and inference wall-clock time on HumanEval with Llama-3.1-8B-Instruct. All experiments are measured on a single NVIDIA GeForce RTX 5090.}
    \label{tab:wallclock_time}
    \begin{tabular}{lccc}
        \toprule
        Method & Training Time (min) & Inference Time (min) & Accuracy \\
        \midrule
        ADAS & 200 & 91 & 59.5 \\
        AFlow & 97 & 66 & 65.1 \\
        MaAS & \best{31} & 22 & 62.7 \\
        \method{} & 39 & \best{13} & 69.6 \\
        \bottomrule
    \end{tabular}
\end{table}

\subsubsection{Detailed Computational Overhead}
\label{app:computation_overhead}

We further profile the additional computation introduced by the IOM and CDM modules during \method{} training. Table~\ref{tab:computation_overhead} reports the time spent on IOM, CDM, and embedding computation on HumanEval. The IOM time includes both encoding the operator outputs and computing the information-guided reward, while the embedding time denotes the total time spent invoking the embedding model throughout training.

The IOM and CDM computations require only 1.81 and 0.74 minutes, respectively, compared with 39 minutes of total training time. Together, these two modules account for approximately 6.5\% of the overall training wall-clock time, indicating that the intermediate reward computation introduces limited additional overhead.

\begin{table}[!t]
    \centering
    \small
    \setlength{\tabcolsep}{7pt}
    \renewcommand{\arraystretch}{1.12}
    \caption{Detailed training-time overhead of \method{} on HumanEval. IOM Time includes the embedding computation required for the IOM reward, while Embedding Time denotes the total embedding-model computation throughout training.}
    \label{tab:computation_overhead}
    \begin{tabular}{lcccc}
        \toprule
        Method & Total Training & IOM Time & CDM Time & Embedding Time \\
        & Time (min) & (min) & (min) & (min) \\
        \midrule
        \method{} & 39 & 1.81 & 0.74 & 2.89 \\
        \bottomrule
    \end{tabular}
\end{table}




\FloatBarrier

\section{OPERATORS}
\subsection{OPERATORS Register}
\begin{center}
\begin{tcolorbox}[
  breakable,
  colback=customblue_back,
  colframe=customblue,
  coltitle=white,
  fonttitle=\bfseries\small\scshape,
  title=Operator Registry,
  arc=4mm,
  boxrule=0.8pt,
  width=0.95\textwidth,
]
\lstset{style=prismjson}
\begin{lstlisting}
{
  "Generate": {
    "description": "Generates anything based on customized input and instruction.",
    "interface": "generate(input: str, instruction: str) -> dict with key 'response' of type str"
  },
  "GenerateCoT": {
    "description": "Generates anything using a chain-of-thought approach, providing step-by-step reasoning before producing the final solution. If you need to solve mathematical or difficult reasoning problems, you can use this operator.",
    "interface": "generate_cot(input: str, instruction: str) -> dict with key 'response' of type str"
  },
  "MultiGenerateCoT": {
    "description": "Generates multiple solutions using diverse chain-of-thought reasoning processes to increase solution variety and robustness.",
    "interface": "multi_generate_cot(input: str, instruction: str) -> dict with key 'response' of type List[str]"
  },
  "VoteEnsemble": {
    "description": "Selects the best solution from multiple candidates using voting, semantic equivalence, and reasoning-based judgment to improve robustness and quality.",
    "interface": "vote_ensemble(solutions: List[str], problem: str) -> dict with key 'response' of type str"
  },
  "Decomposer": {
    "description": "Decomposes a complex problem into structured subtasks with potential dependencies and execution strategies.",
    "interface": "decompose(problem: str) -> dict with key 'subtasks' of type List[dict]"
  },
  "Aggregator": {
    "description": "Aggregates and fuses multiple inputs, intermediate results, or task outputs into a unified and coherent representation.",
    "interface": "aggregate(inputs: List[str], instruction: str) -> dict with key 'response' of type str"
  },
  "Programmer": {
    "description": "Automatically writes, executes Python code, and returns the solution based on the provided problem description and analysis. The `output` only contains the final answer. If you want to see the detailed solution process, it's recommended to retrieve the `code`.",
    "interface": "programmer(problem: str, analysis: str = 'None') -> dict with keys 'code' and 'output' of type str"
  },
  "Test": {
    "description": "Tests the solution using public test cases. If the solution fails, it reflects on the errors and attempts to modify the solution. Returns True and the solution if all tests pass after modifications. Returns False and the current solution if it still fails after modifications.",
    "interface": "test(problem: str, solution: str, entry_point: str) -> dict with key 'result' of type bool and key 'solution' of type str"
  },
  "SelfRefine": {
    "description": "Refines the generated solution by analyzing errors or suboptimal aspects and making iterative improvements to enhance correctness and efficiency.",
    "interface": "self_refine(problem: str, solution: str) -> dict with key 'response' of type str"
  },
  "EarlyStop": {
    "description": "This issue has been correctly and effectively resolved. In order to prevent further activation of operators from wasting resources, we will immediately stop answering this question. Immediately terminates the workflow when selected, preventing any further operators from being invoked.",
    "interface": "early_stop() -> dict with key 'terminate' of type bool"
  }
}
\end{lstlisting}
\end{tcolorbox}
\end{center}

\subsection{OPERATORS Implementation}
\begin{center}
\begin{tcolorbox}[
  breakable,
  colback=customblue_back,
  colframe=customblue,
  coltitle=white,
  fonttitle=\bfseries\small\scshape,
  title=Operators,
  arc=4mm,
  boxrule=0.8pt,
  width=0.95\textwidth,
]
\lstset{style=prismpython}
\begin{lstlisting}
class Operator:
    def __init__(self, llm: LLM, name: str):
        self.name = name
        self.llm = llm
    def __call__(self, *args, **kwargs):
        raise NotImplementedError
    async def _fill_node(self, op_class, prompt, mode=None, **extra_kwargs):
        fill_kwargs = {"context": prompt, "llm": self.llm}
        if mode:
            fill_kwargs["mode"] = mode
        fill_kwargs.update(extra_kwargs)
        node = await ActionNode.from_pydantic(op_class).fill(**fill_kwargs)
        result = node.instruct_content.model_dump()
        if any(str(v or "").strip() for v in result.values()):
            return result
        raw_text = await self.llm.aask(prompt)
        raw_text = str(raw_text or "").strip()
        if not raw_text:
            return result
        if "reflection_and_solution" in result:
            result["reflection_and_solution"] = raw_text
        elif "response" in result:
            result["response"] = raw_text
        return result
    async def _ask_text(self, prompt: str) -> str:
        response = await self.llm.aask(prompt)
        return response

class Generate(Operator):
    def __init__(self, llm: LLM, name: str = "Generate"):
        super().__init__(llm, name)
    async def __call__(self, problem, entry_point, instruction):
        prompt = instruction + problem
        return await self._fill_node(GenerateOp, prompt, mode="code_fill", function_name=entry_point)

class GenerateCoT(Operator):
    def __init__(self, llm: LLM, name: str = "GenerateCoT"):
        super().__init__(llm, name)
    async def __call__(self, problem, entry_point, subtask="", current_solution=""):
        prompt = GENERATE_COT_PROMPT.format(
            problem=problem,
            entry_point=entry_point,
            subtask_block=f"Subtask focus: {subtask}\n" if subtask else "",
            current_solution_block=(
                f"Current draft solution:\n{current_solution}\n\nRevise or improve it.\n"
                if current_solution else ""
            ),
        )
        return await self._fill_node(GenerateOp, prompt, mode="code_fill", function_name=entry_point)

class MultiGenerateCoT(Operator):
    def __init__(self, llm: LLM, name: str = "MultiGenerateCoT"):
        super().__init__(llm, name)
    async def __call__(self, problem, entry_point, current_solution=""):
        candidates = []
        seen = set()
        for variant in MULTI_GENERATE_COT_VARIANTS:
            prompt = GENERATE_COT_PROMPT.format(
                problem=problem,
                entry_point=entry_point,
                subtask_block=f"Additional guidance: {variant}\n",
                current_solution_block=(
                    f"Current draft solution:\n{current_solution}\n\nProduce an alternative improved implementation.\n"
                    if current_solution else ""
                ),
            )
            response = await self._fill_node(GenerateOp, prompt, mode="code_fill", function_name=entry_point)
            candidate = str(response.get("response", "")).strip()
            if candidate and candidate not in seen:
                seen.add(candidate)
                candidates.append(candidate)
        if not candidates:
            fallback = await GenerateCoT(self.llm).__call__(
                problem=problem,
                entry_point=entry_point,
                current_solution=current_solution,
            )
            if str(fallback.get("response", "")).strip():
                candidates.append(fallback["response"])
        return {"response": candidates}

class VoteEnsemble(Operator):
    def __init__(self, llm: LLM, name: str = "VoteEnsemble"):
        super().__init__(llm, name)
    async def __call__(self, problem, entry_point, parent_solutions: List[str]):
        formatted_solutions = []
        for index, solution in enumerate(parent_solutions, start=1):
            formatted_solutions.append(f"Candidate {index}:\n{solution}")
        prompt = VOTE_ENSEMBLE_PROMPT.format(
            problem=problem,
            entry_point=entry_point,
            parent_solutions="\n\n".join(formatted_solutions) if formatted_solutions else "<empty>",
        )
        return await self._fill_node(GenerateOp, prompt, mode="code_fill", function_name=entry_point)

class Decomposer(Operator):
    def __init__(self, llm: LLM, name: str = "Decomposer"):
        super().__init__(llm, name)
    @staticmethod
    def _parse_subtasks(raw_text: str) -> List[str]:
        lines = [line.strip() for line in raw_text.splitlines() if line.strip()]
        subtasks = []
        for line in lines:
            cleaned = re.sub(r"^(?:[-*]|\d+[.)])\\s*", "", line).strip()
            if cleaned:
                subtasks.append(cleaned)
        deduped = []
        seen = set()
        for task in subtasks:
            key = task.lower()
            if key not in seen:
                seen.add(key)
                deduped.append(task)
        return deduped[:4]
    async def __call__(self, problem, entry_point, current_solution=""):
        prompt = DECOMPOSER_PROMPT.format(
            problem=problem,
            entry_point=entry_point,
            current_solution=current_solution or "<empty>",
        )
        raw_response = await self._ask_text(prompt)
        subtasks = self._parse_subtasks(raw_response)
        if not subtasks:
            subtasks = [
                "Derive a correct algorithm and invariants for the target function.",
                "Stress test edge cases and boundary conditions.",
            ]
        return {"response": raw_response, "subtasks": subtasks}

class Aggregator(Operator):
    def __init__(self, llm: LLM, name: str = "Aggregator"):
        super().__init__(llm, name)
    async def __call__(self, problem, entry_point, parent_solutions: List[str]):
        formatted_solutions = []
        for index, solution in enumerate(parent_solutions, start=1):
            formatted_solutions.append(f"Candidate {index}:\n{solution}")
        prompt = AGGREGATOR_PROMPT.format(
            problem=problem,
            entry_point=entry_point,
            parent_solutions="\n\n".join(formatted_solutions) if formatted_solutions else "<empty>",
        )
        return await self._fill_node(GenerateOp, prompt, mode="code_fill", function_name=entry_point)

class Programmer(Operator):
    def __init__(self, llm: LLM, name: str = "Programmer"):
        super().__init__(llm, name)
    async def __call__(self, problem, entry_point):
        prompt = PROGRAMMER_PROMPT.format(problem=problem, entry_point=entry_point)
        return await self._fill_node(GenerateOp, prompt, mode="code_fill", function_name=entry_point)

class Test(Operator):
    INTERNAL_TEST_TIMEOUT_SECONDS = int(os.getenv("MAAS_HUMANEVAL_INTERNAL_TEST_TIMEOUT_SECONDS", "10"))
    def __init__(self, llm: LLM, name: str = "Test"):
        super().__init__(llm, name)
    @staticmethod
    def _exec_code_impl(solution, entry_point):
        test_cases = extract_test_cases_from_jsonl(entry_point, dataset="HumanEval")
        fail_cases = []
        for test_case in test_cases:
            test_code = test_case_2_test_function(solution, test_case, entry_point)
            try:
                exec(test_code, globals())
            except AssertionError as e:
                exc_type, exc_value, exc_traceback = sys.exc_info()
                tb_str = traceback.format_exception(exc_type, exc_value, exc_traceback)
                error_information = {
                    "test_fail_case": {
                        "test_case": test_case,
                        "error_type": "AssertionError",
                        "error_message": str(e),
                        "traceback": tb_str,
                    }
                }
                fail_cases.append(error_information)
            except Exception as e:
                return {"exec_fail_case": str(e)}
        return fail_cases if fail_cases else "no error"
    def exec_code(self, solution, entry_point):
        result_queue = mp.Queue()
        def target(queue, code_solution, code_entry_point):
            try:
                queue.put(("result", self._exec_code_impl(code_solution, code_entry_point)))
            except Exception as e:
                queue.put(("error", str(e)))
        process = mp.Process(target=target, args=(result_queue, solution, entry_point), daemon=True)
        process.start()
        process.join(self.INTERNAL_TEST_TIMEOUT_SECONDS)
        try:
            if process.is_alive():
                process.terminate()
                process.join(timeout=1)
                return {
                    "exec_fail_case": (
                        f"Internal HumanEval public-test execution timed out after "
                        f"{self.INTERNAL_TEST_TIMEOUT_SECONDS} seconds"
                    )
                }
            if result_queue.empty():
                return {"exec_fail_case": "Internal HumanEval public-test execution returned no result"}
            status, payload = result_queue.get_nowait()
            if status == "error":
                return {"exec_fail_case": payload}
            return payload
        finally:
            result_queue.close()
            result_queue.join_thread()
    async def __call__(self, problem, solution, entry_point, test_loop: int = 1):
        if not str(solution or "").strip():
            return {"result": False, "solution": ""}
        for _ in range(test_loop):
            result = self.exec_code(solution, entry_point)
            if result == "no error":
                return {"result": True, "solution": solution}
            if "exec_fail_case" in result:
                exec_result = result["exec_fail_case"]
                prompt = REFLECTION_ON_PUBLIC_TEST_PROMPT.format(
                    problem=problem,
                    solution=solution,
                    exec_pass=f"executed unsuccessfully, error:\n{exec_result}",
                    test_fail="executed unsuccessfully",
                )
            else:
                prompt = REFLECTION_ON_PUBLIC_TEST_PROMPT.format(
                    problem=problem,
                    solution=solution,
                    exec_pass="executed successfully",
                    test_fail=result,
                )
            response = await self._fill_node(ReflectionTestOp, prompt, mode="code_fill")
            solution = response["reflection_and_solution"]
        result = self.exec_code(solution, entry_point)
        return {"result": result == "no error", "solution": solution}
        
class SelfRefine(Operator):
    def __init__(self, llm: LLM, name: str = "SelfRefine"):
        super().__init__(llm, name)
    async def __call__(self, problem, solution):
        prompt = SELFREFINE_PROMPT.format(problem=problem, solution=solution)
        return await self._fill_node(SelfRefineOp, prompt, mode="code_fill")
\end{lstlisting}
\end{tcolorbox}
\end{center}


%% file: paper_sections/checklist.tex
\section*{NeurIPS Paper Checklist}

The checklist is designed to encourage best practices for responsible machine learning research, addressing issues of reproducibility, transparency, research ethics, and societal impact. Do not remove the checklist: {\bf The papers not including the checklist will be desk rejected.} The checklist should follow the references and follow the (optional) supplemental material.  The checklist does NOT count towards the page
limit. 

Please read the checklist guidelines carefully for information on how to answer these questions. For each question in the checklist:
\begin{itemize}
    \item You should answer \answerYes{}, \answerNo{}, or \answerNA{}.
    \item \answerNA{} means either that the question is Not Applicable for that particular paper or the relevant information is Not Available.
    \item Please provide a short (1--2 sentence) justification right after your answer (even for \answerNA). 
\end{itemize}

{\bf The checklist answers are an integral part of your paper submission.} They are visible to the reviewers, area chairs, senior area chairs, and ethics reviewers. You will also be asked to include it (after eventual revisions) with the final version of your paper, and its final version will be published with the paper.

The reviewers of your paper will be asked to use the checklist as one of the factors in their evaluation. While \answerYes{} is generally preferable to \answerNo{}, it is perfectly acceptable to answer \answerNo{} provided a proper justification is given (e.g., error bars are not reported because it would be too computationally expensive'' or ``we were unable to find the license for the dataset we used''). In general, answering \answerNo{} or \answerNA{} is not grounds for rejection. While the questions are phrased in a binary way, we acknowledge that the true answer is often more nuanced, so please just use your best judgment and write a justification to elaborate. All supporting evidence can appear either in the main paper or the supplemental material, provided in appendix. If you answer \answerYes{} to a question, in the justification please point to the section(s) where related material for the question can be found.

IMPORTANT, please:
\begin{itemize}
    \item {\bf Delete this instruction block, but keep the section heading ``NeurIPS Paper Checklist"},
    \item  {\bf Keep the checklist subsection headings, questions/answers and guidelines below.}
    \item {\bf Do not modify the questions and only use the provided macros for your answers}.
\end{itemize}


\begin{enumerate}

\item {\bf Claims}
    \item[] Question: Do the main claims made in the abstract and introduction accurately reflect the paper's contributions and scope?
    \item[] Answer: \answerYes{} 
    \item[] Justification: We have precisely described them in the abstract and introduction.

\item {\bf Limitations}
    \item[] Question: Does the paper discuss the limitations of the work performed by the authors?
    \item[] Answer: \answerYes{} 
    \item[] Justification: We discuss the main limitations and computational efficiency in the experimental section and supplementary appendix. 

\item {\bf Theory assumptions and proofs}
    \item[] Question: For each theoretical result, does the paper provide the full set of assumptions and a complete (and correct) proof?
    \item[] Answer: \answerNA{} 
    \item[] Justification: This is not a theory paper. 

    \item {\bf Experimental result reproducibility}
    \item[] Question: Does the paper fully disclose all the information needed to reproduce the main experimental results of the paper to the extent that it affects the main claims and/or conclusions of the paper (regardless of whether the code and data are provided or not)?
    \item[] Answer: \answerYes{} 
    \item[] Justification: The source code are provided in the appendix. The experimental settings are described in the experimental section, and the appendix. 

\item {\bf Open access to data and code}
    \item[] Question: Does the paper provide open access to the data and code, with sufficient instructions to faithfully reproduce the main experimental results, as described in supplemental material?
    \item[] Answer: \answerYes{} 
    \item[] Justification: The source code are included in the supplementary materials. All the data and the source code will be made open-access upon publication.

\item {\bf Experimental setting/details}
    \item[] Question: Does the paper specify all the training and test details (e.g., data splits, hyperparameters, how they were chosen, type of optimizer) necessary to understand the results?
    \item[] Answer: \answerYes{} 
    \item[] Justification: The details are provided in the experimental section and the appendix.

\item {\bf Experiment statistical significance}
    \item[] Question: Does the paper report error bars suitably and correctly defined or other appropriate information about the statistical significance of the experiments?
    \item[] Answer: \answerYes{} 
    \item[] Justification: The variance are reported along with the mean value.

\item {\bf Experiments compute resources}
    \item[] Question: For each experiment, does the paper provide sufficient information on the computer resources (type of compute workers, memory, time of execution) needed to reproduce the experiments?
    \item[] Answer: \answerYes{} 
    \item[] Justification: The experiments are conducted in a computer clusters with multiple Nvidia H100 (80G) GPUs. 
    
\item {\bf Code of ethics}
    \item[] Question: Does the research conducted in the paper conform, in every respect, with the NeurIPS Code of Ethics \url{https://neurips.cc/public/EthicsGuidelines}?
    \item[] Answer: \answerYes{} 
    \item[] Justification: We follow the NeurIPS Code of Ethics. No human subject, and private datasets are involed. We use public available datasets that are shared to the research community. 

\item {\bf Broader impacts}
    \item[] Question: Does the paper discuss both potential positive societal impacts and negative societal impacts of the work performed?
    \item[] Answer: \answerYes{} 
    \item[] Justification: Our work is foundational research and not tied to particular applications.
    
\item {\bf Safeguards}
    \item[] Question: Does the paper describe safeguards that have been put in place for responsible release of data or models that have a high risk for misuse (e.g., pre-trained language models, image generators, or scraped datasets)?
    \item[] Answer: \answerNA{} 
    \item[] Justification: Such paper poses no such risks.

\item {\bf Licenses for existing assets}
    \item[] Question: Are the creators or original owners of assets (e.g., code, data, models), used in the paper, properly credited and are the license and terms of use explicitly mentioned and properly respected?
    \item[] Answer: \answerYes{} 
    \item[] Justification: We have cited the original paper that produced the code and the dataset.

\item {\bf New assets}
    \item[] Question: Are new assets introduced in the paper well documented and is the documentation provided alongside the assets?
    \item[] Answer: \answerYes{} 
    \item[] Justification: We will release the code of this work. 

\item {\bf Crowdsourcing and research with human subjects}
    \item[] Question: For crowdsourcing experiments and research with human subjects, does the paper include the full text of instructions given to participants and screenshots, if applicable, as well as details about compensation (if any)? 
    \item[] Answer: \answerNA{} 
    \item[] Justification: No human subjects were involved.

\item {\bf Institutional review board (IRB) approvals or equivalent for research with human subjects}
    \item[] Question: Does the paper describe potential risks incurred by study participants, whether such risks were disclosed to the subjects, and whether Institutional Review Board (IRB) approvals (or an equivalent approval/review based on the requirements of your country or institution) were obtained?
    \item[] Answer: \answerNA{} 
    \item[] Justification: Not applicable.

\item {\bf Declaration of LLM usage}
    \item[] Question: Does the paper describe the usage of LLMs if it is an important, original, or non-standard component of the core methods in this research? Note that if the LLM is used only for writing, editing, or formatting purposes and does \emph{not} impact the core methodology, scientific rigor, or originality of the research, declaration is not required.
    \item[] Answer: \answerYes{} 
    \item[] Justification: We use an LLM to polish the paper. In this work, we study LLM-based agentic workflows. LLMs are used as the base models invoked by our proposed workflow method. An LLM is a necessary component of our work.

\end{enumerate}